\documentclass[preprint,amsmath,amssymb,aps]{revtex4-1}
\usepackage{CJK}
\usepackage{graphicx}
\usepackage{color}

\usepackage{dcolumn} 
\usepackage{bm}     
\usepackage{amsfonts}
\usepackage[bookmarks=false,colorlinks,citecolor=red]{hyperref}
\usepackage{amsmath}
\newcommand{\angstrom}{\textup{\AA}}
\usepackage[version=4]{mhchem}
\usepackage{siunitx}

\include{MyCommand}

\newif\ifshowcomments\showcommentstrue

\begin{document}

\title{Many-body Electronic Structure of NdNiO$_2$ and CaCuO$_2$}
\author{Jonathan Karp$^1$}
\email{jk3986@columbia.edu}
\author{Antia S. Botana$^2$, Michael R. Norman$^3$, Hyowon Park$^{3,4}$, Manuel Zingl$^5$, Andrew Millis$^{5,6}$}
\affiliation{
$^1$  Department of Applied Physics and Applied Math, Columbia University, New York, NY, 10027\\ 
$^2$ Department of Physics, Arizona State University, Tempe, AZ  85287\\ 
$^3$ Materials Science Division, Argonne National Laboratory, Argonne, IL 60439\\
$^4$ Department of Physics, University of Illinois at Chicago, Chicago, IL, 60607\\
$^5$ Center for Computational Quantum Physics, Flatiron Institute, 162 5th Avenue, New York, NY, 10010 \\
$^6$  Department of Physics, Columbia University, New York, NY, 10027}

\begin{abstract}
The demonstration of superconductivity in nickelate analogues of high $T_c$ cuprates provides new perspectives on the physics of correlated electron materials. The degree to which the nickelate electronic structure is similar to that of cuprates is an important open question. This paper presents results of a comparative study of the many-body electronic structure and theoretical phase diagram of the  isostructural materials CaCuO$_2$ and NdNiO$_2$. Both NdNiO$_2$ and CaCuO$_2$ are found to be charge transfer materials. Important differences include the proximity of the oxygen $2p$ bands to the Fermi level, the bandwidth of the transition metal-derived $3d$ bands, and the presence, in NdNiO$_2$, of both Nd-derived $5d$ states crossing  the Fermi level and a van Hove singularity that crosses the Fermi level as the out of plane momentum is varied. The low energy physics of NdNiO$_2$ is found to be that of a single Ni-derived correlated band, with additional accompanying weakly correlated bands of Nd-derived states that dope the Ni-derived band. The effective correlation strength of the Ni-derived $d$-band crossing the Fermi level in NdNiO$_2$ is found to be greater than that of the Cu-derived $d$-band in CaCuO$_2$, but the predicted magnetic transition temperature of NdNiO$_2$  is substantially lower than that of CaCuO$_2$ because of the smaller bandwidth.
\end{abstract}

\maketitle

{\bf Introduction:} The remarkable physics of layered copper-oxide materials~\cite{Anderson87,Pickett89,Keimer15,Proust19}, including high transition temperature superconductivity for carrier concentrations not too far from optimal doping,  electronic pseudogaps, and various forms of long- and short-range order, has challenged researchers over the  more than thirty years since the discovery of superconductivity in La$_{2-x}$Ba$_x$CuO$_4$~\cite{Bednorz86}.  Many cuprate superconductors are known; all share the structural motif of  CuO$_2$ planes weakly coupled in the third dimension and the electronic motif of an approximately $d^9$ electronic configuration for Cu, but differ in other details.  A  key  question is whether the novel physics of these materials can be essentially understood in a one-band model with  strong local correlations~\cite{Anderson87}  or whether other physics  is important~\cite{Emery87,Littlewood89}.

It was recognized early on  that perspective on this issue could be gained from the analysis of materials with similar features  but with a different local chemistry. Ni, which is adjacent to Cu on the periodic table, has been of particular interest in this regard~\cite{Anisimov99,Lee04,Liu14,Botana18}. Because Ni has one fewer proton than Cu, the $d$-electron count for Ni is typically expected to be lower than for Cu, but with appropriate chemistry a configuration close to  Ni $d^9$, with one hole in the $d_{x^2-y^2}$ orbital, might be achieved.  Proposals to use artificial superlattices~\cite{Chaloupka08} have yet to yield such a configuration, but recently, trilayer systems with  a formal valence  of Ni$^{4/3+}$ ($d^{8.67}$) and $d_{x^2-y^2}$ holes were synthesized~\cite{Zhang2017}. One of these materials (Pr$_4$Ni$_3$O$_8$) is metallic but not superconducting, likely because the carrier concentration corresponds to a doping far  beyond the optimal doping of typical cuprate superconductors. 

\vspace{0.3cm}
NdNiO$_2$ is isostructural to the ``infinite layer'' cuprate CaCuO$_2$ and has a formal Ni $d^9$ valence. The recent discovery~\cite{Hwang2019} that NdNiO$_2$ can be synthesized,  hole doped, and made superconducting has created intense excitement. There has been an outpouring of theoretical interest, with DFT~\cite{botana2019, nomura2019, hepting2019, wu2019, sakakibara2019, gao2019electronic, zhang2019effective, jiang2019electronic, hirayama2019materials, lechermann2019late, gu2019, ryee2019induced, si2019topotactic, choi2019role, liu2019electronic}, DMFT~\cite{gu2019, ryee2019induced, werner2019nickelate, si2019topotactic, lechermann2019late}, and model-system~\cite{wu2019, sakakibara2019, jiang2019, Hirsch2019, zhang2019selfdoped, zhang2019effective, zhang2019kondo, hu2019twoband} studies of the material. These papers have come to a variety of conclusions about the important low energy physics of the nickelate materials and its relation to the low energy physics of the cuprates. In this paper we aim to  clarify some of the issues.

Because transition metal oxides such as cuprates and nickelates are relatively ionic,  the basic electronic structure problem  may be posed in terms of corrections to a  ``formal valence'' analysis of isolated ion electron affinities  and ionization energies. The formal valence analysis provides an estimate of the average occupancies of the relevant orbitals including the transition metal $d$ and oxygen $p$ states. Hybridization in the solid leads to fluctuations about these average occupancies; the fluctuations are constrained by the strong local correlations of the transition metal $d$ states, and the implications for the low energy physics  of the resulting  correlated hybridization problem  need to be determined.

\begin{figure*}[t]
\centering
\includegraphics[width = 0.97\textwidth]{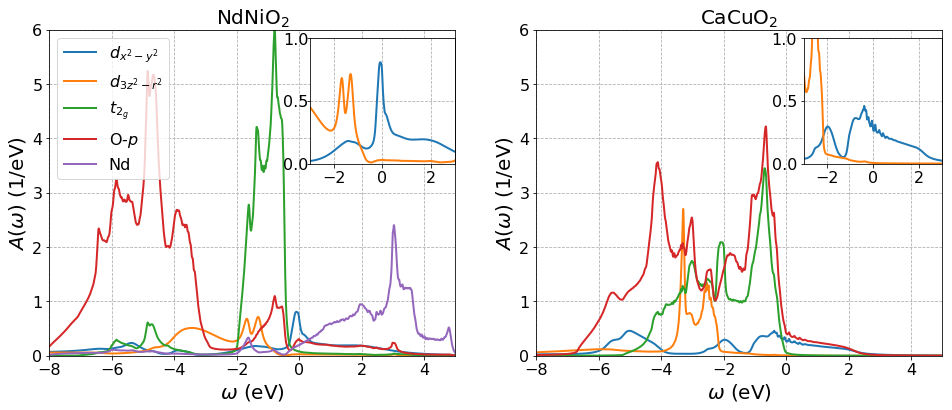}
\caption{Main panels: orbitally resolved wide energy window DFT+DMFT spectral function summed over spin computed for  NdNiO$_2$ (left panel) and CaCuO$_2$ (right panel) in the paramagnetic phase, treating the transition metal $d_{x^2-y^2}$ and $d_{3z^2-r^2}$ orbitals as correlated,  with on-site interaction $U$ = \SI{7}{eV} and  $J=\SI{0.7}{eV}$ at temperature $T = \SI{290}{K}$. Insets: expanded view of near chemical potential region showing only $d_{x^2-y^2}$ and $d_{3z^2-r^2}$ contributions to the spectral function. The zero of energy is defined to be the chemical potential.}
\label{ComparativeDOS}
\end{figure*}

For cuprates such as  CaCuO$_2$, formal valence and crystal symmetry considerations lead to the configuration Ca$^{2+}$ Cu$^{2+}$ ($d^9$) O$^{2-}$ ($p^6$)  with the Cu hole in the $d_{x^2-y^2}$ orbital. The Ca-$d$ states are very high in energy and weakly coupled to the Cu-$d$ and O-$p$ states of relevance at lower energies, so Ca$^{2+}$ may be regarded as electrically inert. The two potentially relevant charge fluctuation processes are transfer of an electron from one Cu to another  ($2d^9\rightarrow d^8d^{10}$) and from an O to a Cu $(d^9\rightarrow d^{10}\underbar{L}$). The energy cost of the former process defines the Hubbard parameter $U=E[d^8d^{10}]-2E[d^9]$, while the energy cost of the latter defines the charge transfer parameter $\Delta=E[d^{10}\underbar{L}]-E[d^9]$. In cuprates it is believed that $U\sim 7-\SI{8}{eV} \gg \Delta \lesssim \SI{3}{eV}$~\cite{Eskes88} so the $d^8$ configuration is essentially irrelevant: the low energy physics may be discussed in the manifold of states defined by the $d^9$ and $d^{10}\underbar{L}$ configurations. 

For NdNiO$_2$  the formal valence considerations are similar, implying an average electronic configuration of  Nd$^{3+}$ Ni$^{1+}$ ($d^9$) O$^{2-}$ ($p^6$). However, the relative importance of the Ni $d^8$ configuration vis a vis $d^{10}\underbar{L}$ is not yet determined. Further, Nd$^{2+}$ is low enough in energy that there is some charge transfer to Nd, meaning that at least the $d^9\underbar{L}$ configuration, and perhaps also the $d^8$, needs to be considered even in the stoichiometric material. If $d^8$  is important, one may ask whether the relevant $d^8$ state is high spin (implying one hole in the $d_{x^2-y^2}$ and one in the $d_{3z^2-r^2}$), as proposed or implied by some calculations~\cite{Liu14,botana2019,lechermann2019late,zhang2019kondo,jiang2019,hu2019twoband}, or low spin (both holes in the $d_{x^2-y^2}$ orbitals). 

In this paper, we address these issues via a comparative study of the many-body electronic structure of NdNiO$_2$ and the isostructural cuprate CaCuO$_2$. We use the density functional theory plus dynamical mean-field theory method (DFT+DMFT)~\cite{Georges1996,Georges04,Kotliar06,Held06}, building on previous DFT ~\cite{Lee04,botana2019} and DFT+DMFT~\cite{gu2019, ryee2019induced, werner2019nickelate, si2019topotactic, lechermann2019late} studies to determine many-body densities of states, orbital occupancies, phase boundaries, and mass enhancements.

{\bf Methods:} Our study, in common with most theoretical studies of many-body electronic structure of solids, is formulated in a subspace of the full electronic Hilbert space that is constructed from single particle states obtained from a relatively inexpensive mean field-type calculation. We employ DFT (Quantum Espresso~\cite{QE} with the PBE-GGA exchange-correlation functional~\cite{PBE}); our DFT results are in agreement with prior literature~\cite{wu2019, botana2019, nomura2019, hepting2019, sakakibara2019, gao2019electronic, zhang2019effective, jiang2019, hirayama2019materials, lechermann2019late}. We then use Wannier90~\cite{wannier90_v3} to construct a basis of maximally-localized Wannier functions~\cite{MLWF1, MLWF2} spanning an energy window of interest. We employ two energy windows: One is a wide energy window that includes the full transition metal $d$-manifold, the O-$p$ states, and, in  NdNiO$_2$,  the relevant Nd states. The second is a narrow energy window that includes only the ``frontier orbitals'' that cross the chemical potential. The wide energy window enables a natural connection to  formal valence considerations and high energy spectroscopies and helps define the frontier orbital model, while the frontier orbital window is better suited to discussions of the low energy physics.  In both cases the resulting  Wannier bands  reproduce the DFT bands very well.  Details, including  lattice constants, $k$-mesh and energy cutoffs used in the DFT calculations,  specification of the energy windows in the Wannierization, visualization of some of the Wannier functions, and comparison of Wannier bands to the DFT bands, are provided in the Supplementary Material. 

The Wannierization of the Nd-derived $5d$ states in NdNiO$_2$ has been the subject of  discussion in the literature~\cite{nomura2019, hepting2019, sakakibara2019, hirayama2019materials}. We use selective localization methods~\cite{Wang14} as implemented in Wannier90~\cite{wannier90_v3} to localize only the Ni-$d$ Wannier functions, which are also constrained to be centered on the Ni sites. The resulting fit yields not only physically reasonable Ni-d Wannier functions but also two Nd-centered orbitals, one of $xy$ and one of $3z^2-r^2$ symmetry.

\renewcommand{\arraystretch}{1.8}
\begin{table*}[t]
\caption{Left portion: occurrence probabilities $P$[config] of transition metal d configurations ($e_g$ manifold). Right portion: occupancies N$_{\text{state}}$ of transition metal $d_{x^2-y^2}$, O-$p_\sigma$ and Nd-5$d$ orbitals (summed over spin). }
\begin{center}
\begin{tabular}{|c|c|c|c|c|c|c|c|c|c|}
\hline
&~~~P$\left[d^{10}\right]$~~~&P$\left[d^{9}_{x^2-y^2}\right]$&P$\left[d^{9}_{3z^2-r^2}\right]$&~~~P$\left[d^8\right]~~~$& &~N$_{d_{x^2-y^2}}$~&~N$_{d_{3z^2-r^2}}$~&~~N$_{\text{O-}{p_\sigma}}$~~&~~~~N$_{\text{Nd}}$~~~~\\
\hline 
~~NdNiO$_2$~~&0.26&0.65&0.04&0.05&&1.28&1.93&3.54&0.53\\
\hline
~~CaCuO$_2$~~&0.55&0.43&0.00&0.02&&1.54&2.00&3.50& -\\
\hline
\end{tabular}
\end{center}
\label{orbitaloccupancy}
\end{table*}

We define an effective Hamiltonian by projecting the Kohn-Sham states onto the Wannier basis and adding interactions that couple some of the states in the subspace. The interaction terms depend on the energy window and the choice of correlated orbitals, both because the sizes of the relevant orbitals are window-dependent and because the value of $U$ depends on screening which again is affected by the energy window. For our wide energy window calculations we choose the transition metal $d_{x^2-y^2}$ and $d_{3z^2-r^2}$ orbitals as the correlated subspace and use an interaction of Kanamori form~\cite{Kanamori1963} with $U = \SI{7}{eV}$ and $J = \SI{0.7}{eV}$, representative of nickelates~\cite{Nowadnick15} and cuprates~\cite{Eskes88}.
To compensate for the Hartree shifts induced by the added interactions, we include a double counting correction in the form proposed by Held~\cite{Held2007Electronic} (see Supplementary Material for details). For the frontier orbital calculations we correlate only the  $d_{x^2-y^2}$-derived orbitals and consider a range of $U$, noting that recent cRPA calculations suggest a value of $U \approx \SI{3.1}{eV}$~\cite{nomura2019}. We then perform single-site DMFT calculations using the TRIQS software library~\cite{TRIQS, TRIQS/DFTTOOLS} with the continuous-time hybridization expansion solver (CT-HYB)~\cite{TRIQS/CTHYB}. These calculations are ``single shot'' in the sense that the DFT density is not further updated.

{\bf Results:} The main panels of Fig.~\ref{ComparativeDOS} show the many-body density of states projected onto the Wannier orbitals computed in the wide energy window for CaCuO$_2$ (right panel) and NdNiO$_2$ (left panel). In both materials, the removal spectrum ($\omega<0$, with the chemical potential defined as the zero of energy) is dominated by O-$p$ and non-$d_{x^2-y^2}$ states, and the addition spectrum ($\omega>0$) by a transition metal-$d_{x^2-y^2}$ and O-${p}$ hybrid, along with nearly empty Nd-derived bands in NdNiO$_2$. The transition metal $t_{2g}$ and $d_{3z^2-r^2}$ states are essentially filled, although in NdNiO$_2$ close inspection of Fig.~\ref{ComparativeDOS} reveals a tail of these states at $\omega>0$. This tail is a consequence of hybridization with the Nd bands and will be discussed in more detail below. In both materials the near chemical potential transition metal $d_{x^2-y^2}$ spectrum exhibits a van Hove peak, which is at the chemical potential in the Ni material and below in the Cu material, a weak maximum at $\omega\sim \SI{-2}{eV}$, and a gap (due to hybridization with the O-$p$ states) at $\sim 1\text{-}\SI{2}{eV}$ below the chemical potential. The higher lying parts of both the CaCuO$_2$ and NdNiO$_2$ $d_{x^2-y^2}$ and O-$p$ addition spectra are similar, implying a non-negligible mixing of $d_{x^2-y^2}$ and O-${p_\sigma}$ in both materials.

While the broad features of the spectral function are similar, several important differences are immediately evident. In the Ni material, Nd-derived states appear in both the addition and removal spectra.  The resultant charge transfer to the Nd states  dopes the Ni-O complex of states. We see that the centroid of the O-$p$ states is about $\SI{2}{eV}$ lower in NdNiO$_2$ than in CaCuO$_2$. In the Cu compound the oxygen degrees of freedom are very strongly hybridized  with the transition metal degrees of freedom, as can be seen from the close correspondence of the near and above chemical potential O-$p$ and Cu-$d$ addition and removal spectra.  In the Ni material, comparison of the O-$p$ and Ni-$d$  densities of states suggests weaker hybridization with the oxygen degrees of freedom in the immediate vicinity of the chemical potential.

We now consider the electronic states more quantitatively, beginning with the occupation probabilities of the different Wannier orbitals presented in Tab.~\ref{orbitaloccupancy}.  We characterize the (uncorrelated) Nd and O states by their mean occupancies, obtained from an integral over negative energies of the relevant diagonal parts of the on-site portion of the many-body Green's function in the Wannier basis. For the transition metal ions we present the mean occupancies of the $d_{x^2-y^2}$ and $d_{3z^2-r^2}$ states and the occurrence probabilities of different many-body configurations of the correlated orbitals, obtained as in~\cite{Haule_ctqmc}.

For CaCuO$_2$, all of the orbitals except the Cu-$d_{x^2-y^2}$ and O-${p_\sigma}$ are to good approximation fully filled. Measured with respect to the Cu $d^{10}$/O $p^6$ configuration there is one hole, which our calculation finds to be approximately equally shared between the O-$p_\sigma$ and Cu-$d_{x^2-y^2}$ orbitals, ($\sim 50\%$ $d^{10}\underbar{L}$ and $\sim 50\%$ $d^9$). The total weight in the $d^8$ configuration is very small relative to the $d^{10}\underbar{L}$ configuration, consistent with an identification of the cuprates as charge transfer insulators. The occurrence probabilities can be related to features in the spectral function following arguments given by Eskes and Sawatzky~\cite{Eskes88}. Corresponding to the $\sim 50\%$ weight of the $d^{10}\underbar{L}$ configuration, the addition spectrum shows approximately equal weight of O-$p_\sigma$ and Cu-$d_{x^2-y^2}$ character. The feature in the removal spectrum at $\omega\sim \SI{-5}{eV}$ corresponds to the removal of an electron  from a $d^9$ configuration to create $d^8$, while the feature at $\SI{-3}{eV}\lesssim \omega<0$ corresponds to the removal of an electron from $d^{10}\underbar{L}$ to create  $d^9\underbar{L}$. The approximately equal areas of the $\SI{-5}{eV}$ and $\SI{-3}{eV}<\omega < 0$ removal features is consistent with the calculated small proportion of $d^8$ in the ground state.  

Turning now to the Ni material we find that the Nd ions capture about $0.5$ electrons (summed over both relevant Nd orbitals). A similar charge transfer is found in DFT and DFT+U calculations. About $0.3$ of these electrons come from the Ni-$t_{2_g}$, Ni-$d_{3z^2-r^2}$, and O $p_\pi$ and $p_z$ orbitals. The remaining $0.2$ is transferred from the Ni-$d_{x^2-y^2}$ and O-$p_\sigma$ complex, leading to a density of O-$p_\sigma$ holes very much similar to that of the cuprates and $d_{x^2-y^2}$ density about $0.2$ electrons lower. The different ratio of $d_{x^2-y^2}$ and O-$p_\sigma$ holes in the nickelate vs. the cuprate is a reflection of the larger charge transfer energy in the nickelate. Most of this charge transfer is a hybridization effect, arising because occupied orbitals contain an admixture of Nd and higher-lying unoccupied orbitals of Ni-d and O-$p_\sigma$ states. However, as discussed in more detail below in our analysis of the Fermi surface in the narrow energy window case, there is an $\sim 0.1$ electron ``self-doping'' effect arising from  actual charge transfer from primarily Ni-derived bands to primarily Nd-derived bands.

The proportion of $d^8$ in the ground state remains very small (although slightly larger than in the Cu material). This finding is at variance with the results of Ref.~\cite{zhang2019effective} which employs a constrained DFT analysis and concludes that the $d^8$ state is highly relevant in the nickelates. Looking at the spectral functions we see that, consistent with the small $d^8$ fraction, the  relative weight of the low energy $\omega \lesssim \SI{-4}{eV}$ part of the removal spectrum for the Ni material is comparable to that in the near Fermi level part. The peak of the $d^8$ feature is at $\SI{-5.5}{eV}$ in the nickelate and at $\SI{-5}{eV}$ in the cuprate; in both materials the feature exhibits a tail to lower energies, which is longer in the nickelate.  It is interesting that while the admixture of O-$p_\sigma$ and $d_{x^2-y^2}$ is only slightly smaller in the nickelate than in the cuprate, in the nickelates the admixture of O-$p$ orbitals with $t_{2g}$ is substantially smaller. While the wide spread in energy of the oxygen-related removal features makes it difficult to precisely define a charge transfer energy, we may conclude that the DMFT data places the Ni material in the same charge transfer insulator class as the cuprates but with less O-$p$ involvement in the near Fermi-level states and a larger charge transfer gap.

We now consider in more detail the nature of the correlated near chemical potential states in the Ni material.  Several authors have investigated the infinite layer nickelates using the spin density functional plus U  (sDFT+U) method~\cite{Lee04,Liu14,botana2019}, which is in effect the Hartree approximation to the wide energy window  DFT+DMFT method~\cite{park2014computing}, although the use of a spin density functional rather than the simple density functional method is an important difference~\cite{Chen2015Density,Chen16}.  For a range of $U$ including $U=0$ this method finds that the ground state is antiferromagnetic  and  metallic with some hole doping on the majority spin $d_{x^2-y^2}$ orbital and an empty minority spin $d_{x^2-y^2}$ orbital. This state is the Hartree (single determinant) version of the DMFT state reported here. As the magnitude of the on-site interaction is increased beyond a critical value $U_c\approx$ \SI{6}{eV}, the DFT+U method with the FLL double counting scheme~\cite{liechtenstein1995fll}  finds~\cite{Liu14,botana2019} a first-order transition to a state with a fully filled majority spin $d_{x^2-y^2}$ orbital and holes on the $d_{3z^2-r^2}$ orbital in the minority spin channel (the transition is absent if the AMF double counting scheme~\cite{czyzyk1994amf} is used instead). The full occupancy of the majority spin $d_{x^2-y^2}$ state indicates an orbitally selective Mott transition, while the holes in the minority spin channel indicate an admixture of high spin $d^8$ in the ground state, thus identifying the nature of the first-order transition. This transition is not found in the DFT+DMFT calculations at the $U$ values we have studied, but within a DFT+DMFT perspective at $U=\SI{10}{eV}$ Lechermann~\cite{lechermann2019late} finds an orbitally selective state with insulating $d_{x^2-y^2}$ and metallic $d_{3z^2-r^2}$, implying a small but important admixture of d$^8$. Studies of other compounds find that the DFT+U approximation very substantially underestimates the critical $U$ of the low-spin to high-spin transition~\cite{Chen2015Density} and that the spin-polarized DFT functionals overestimate the spin splitting of the $d$ states~\cite{Chen16}. We therefore believe that this aspect of the DFT+U calculations is not relevant to  NdNiO$_2$. However, as an exploration of a theoretically interesting model, a DMFT-based investigation of orbitally selective Mott and high spin-low spin transitions in models with two correlated orbitals coupled to one or more uncorrelated bands would be desirable.

\begin{figure}[t]
\centering
\includegraphics[width= \linewidth]{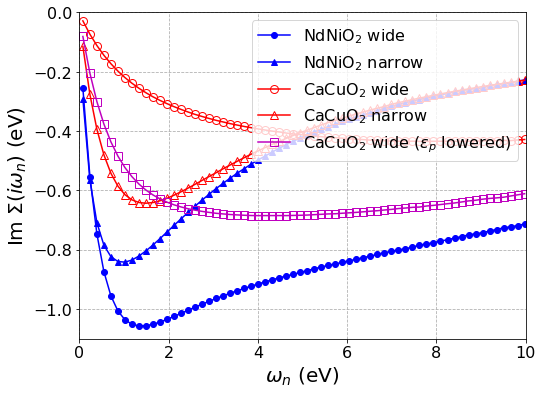}
\caption{Imaginary part of the Matsubara self-energy for the $d_{x^2-y^2}$ orbital  for NdNiO$_2$ (blue, filled symbols) and CaCuO$_2$ (red and magenta, open symbols) at $T = \SI{290}{K}$, for both the wide energy window with $U = \SI{7}{eV}$ (circles)  and the narrow energy window with $U = \SI{3.1}{eV}$ (triangles). Also shown is the wide energy window result for CaCuO$_2$ with $\varepsilon_d - \varepsilon_p$ increased by $\SI{1}{eV}$ (squares, referred to in the legend as ``$\varepsilon_p$ lowered'').}
\label{fig:matsubara_sigma}
\end{figure}

\renewcommand{\arraystretch}{1.8}
\begin{table*}[t]
\begin{tabular}{|c|c|c|c|c|c|c|c|}
\hline
          &~~~~~W~~~~~&~~~~~$t$~~~~~&~~~~~$t^\prime$~~~~~&~~~~~$t^{\prime\prime}$~~~~&~~~~~$t_z$~~~~~& ~~t$^{\text{Ni-Nd}}_{3z^2-r^2}$~~&~~$t^{\text{Ni-Nd}}_{xy}$~~\\ \hline
~~NdNiO$_2$~~& 3.01      & -0.357 & 0.091    & -0.043   & -0.032 & 0.023 & 0.012                            \\ \hline
~~CaCuO$_2$~~& 4.14      & -0.469 & 0.100    & -0.090   & -0.054 &  -   &     -                    \\ \hline
\end{tabular}
\caption{Bandwidth W and near-neighbor hopping parameters (in eV) for the NdNiO$_2$ and CaCuO$_2$ narrow energy window Wannier Hamiltonians. Here, $t$, $t^\prime$ and $t^{\prime\prime}$ are the first, second and third neighbor in-plane hopping parameters (further neighbor hoppings are smaller thus not shown), $t_z$ is the nearest-neighbor interplane hopping parameter, and $t^{\text{Ni-Nd}}$ is the largest  hybridization found between the Ni $d_{x^2-y^2}$ and the two Nd-$5d$ orbitals. }
\label{tab:narrow_params}
\end{table*}

In charge transfer systems the effective interaction strength is determined not only by the $U$ and $J$ parameters but by the double counting correction, which is the subject of some discussion~\cite{Dang14, Haule2015exact} and is one of the significant uncertainties in the DFT+DMFT and DFT+U methodologies. A decreasing of the double counting shifts the $d$-bands up with respect to the $p$-bands, and can thus have the effect of increasing the correlation strength. To show this effect, we have performed additional calculations for CaCuO$_2$ with the O-$p$ orbital energies decreased by $\SI{1}{eV}$, effectively increasing the charge transfer energy $(\varepsilon_d - \varepsilon_p)$. This decreases the admixture of $d^{10}\underbar{L}$ in the ground state and yields a density of states somewhat closer to the one for the nickelate (see Supplementary Material). The occurrence probabilities are  $P\left[d^{10}\right] = 0.42$, $P\left[d^{9}\right] = 0.56$, and $P\left[d^{8}\right] = 0.02$, which are approximately halfway between the results obtained for both compounds using the Held double counting (see Tab.~\ref{orbitaloccupancy}). In NdNiO$_2$, changing the double counting would change both the $p$-$d$ energy splitting and the overlap in energy between Ni-derived and Nd-derived states. The latter effect would change the  doping of the  Ni-derived states, which raises issues of charge self-consistency. The detailed investigation of the effects the double counting correction and full charge self-consistency in NdNiO$_2$ is beyond the scope of this work, but is desirable to be addressed in future works.

Finally, we quantify the strength of correlations via the $d_{x^2-y^2}$ component of the electron self-energy shown in Fig.~\ref{fig:matsubara_sigma}. We see immediately that the self-energy magnitude in the Cu material is much less than in the Ni material. The self-energy corresponding to the calculation with increased charge transfer energy $(\varepsilon_d - \varepsilon_p)$ in CaCuO$_2$ is still somewhat smaller than the self-energy for the nickelate material.
One quantitative metric is the renormalization factor $Z = \left( 1 - \partial \text{Im}\Sigma(i \omega_n)/\partial \omega_n|_{\omega_n \to 0}\right)^{-1}$. For the transition metal $d_{x^2-y^2}$ orbitals these are $Z=0.26$ (Ni) and $Z=0.75$ (Cu) in the wide energy window calculations. In the case of increased $\varepsilon_d - \varepsilon_p$ in the Cu material we find $Z = 0.50$. Note that these are the ``orbital basis'' $Z$ values, determined from the diagonal elements of the projection of the self-energy operator onto the Wannier basis. Also of physical relevance is the ``band basis'' $Z$, describing the states crossing the Fermi level, where the admixture of O-$p_\sigma$ reduces the magnitude of the band basis self-energy. For the Ni material we find that the band basis $Z$ is about $0.33$. For CaCuO$_2$, the band basis $Z$ is roughly $0.83$ and about $0.62$ with increased $\varepsilon_d - \varepsilon_p$ splitting (see Supplementary Material).

\begin{figure*}[t]
\centering
\includegraphics[width = 0.99\textwidth]{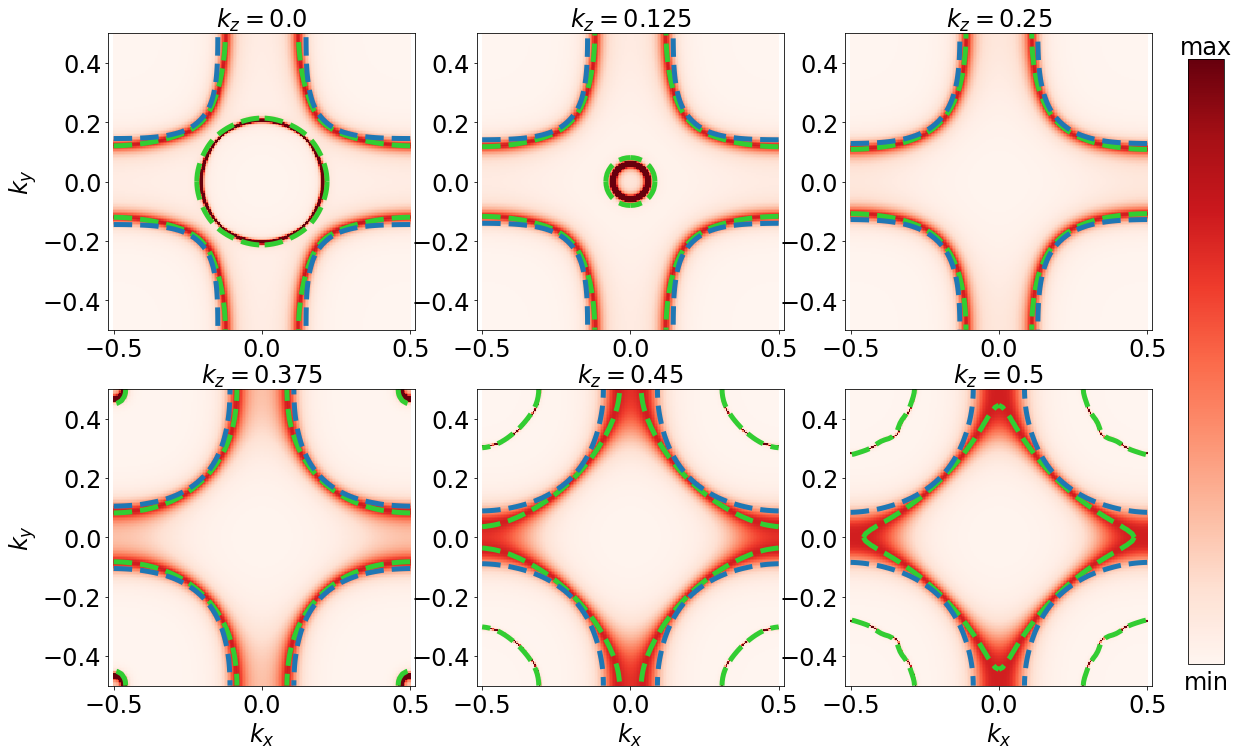}
\caption{ Many-body Fermi surfaces for NdNiO$_2$ defined via a false-color plot of $Tr\left[A(\omega=0)\right]$, shown along with the DFT Fermi surface for NdNiO$_2$ (dashed green line) and CaCuO$_2$ (dashed blue line). Note that for CaCuO$_2$ the DFT and DMFT Fermi surfaces are indistinguishable.  }
\label{fermi_surface}
\end{figure*}

We now investigate the low energy physics by considering a narrow energy window in our Wannier analysis. From the wide energy window analysis, we see that only the transition metal $d_{x^2-y^2}$ and low-lying Nd $5d$-bands are relevant near the Fermi level, so we retain only these orbitals in our Wannierization (see Supplementary Material for details). This yields a one-band (cuprate) or three-band (nickelate) Hamiltonian (see Tab.~\ref{tab:narrow_params} for some of the parameters). We see immediately that the bandwidth $W$ of the cuprate is about $1.4$ times the nickelate bandwidth. This difference is a consequence of the closer proximity of the O-$p$ states to the Fermi level in the cuprate case. Increasing $\varepsilon_d-\varepsilon_p$ by $\SI{1}{eV}$ in the cuprate case yields $W\approx \SI{3.6}{eV}$, smaller than that found with the original $\varepsilon_d - \varepsilon_p$ but still larger than the nickelate bandwidth (Tab.~\ref{tab:narrow_params}). Note that the ratios $t^\prime/t$ and $t^z/t$ are similar although the bandwidths of the two materials are rather different.

To these Wannier Hamiltonian we add an on-site interaction $U$ that correlates the transition metal $d_{x^2-y^2}$-derived states. The subject of the ``correct'' $U$ for the low energy theory is an area of current research, particularly in situations (such as occur in the cuprate materials) where oxygen bands strongly overlap the transition metal bands (for a discussion in the context of pseudocubic perovskite nickelates, see Ref.~\cite{Seth17}). Recent cRPA calculations give $U \approx \SI{3.1}{eV}$ for the Ni material~\cite{nomura2019}.

In Fig.~\ref{fig:matsubara_sigma}, we show the frontier-orbital self-energies for both materials obtained for this $U$ value. The frequency dependence of the self-energy is quite different at high frequencies from that found in the wide energy window calculations, as expected from the further truncation of the Hilbert space in the frontier orbital case, but the behavior in the low frequency regime can be compared.  We obtain $Z = 0.25$ for the Ni-derived band, slightly lower than what we found in the band basis ($Z \sim 0.3$) for the wide energy window calculation. The larger bandwidth of the cuprate material leads to less correlation and only $Z = 0.42$. That is much smaller than the renormalization we obtained in the wide energy window calculations ($Z=0.83$) and still smaller than in the calculation with increased $\varepsilon_d-\varepsilon_p$ difference ($Z=0.62$).

The resulting many-body Fermi surfaces (trace of the many-body spectral function evaluated at $\omega=0$) are shown in Fig.~\ref{fermi_surface} as plots in the space of in-plane momenta for different $k_z$. Due to the smaller self-energy, the many-body Fermi surface is quite sharply defined for the cuprate; it overlaps the non-interacting Fermi surface precisely so only the latter quantity is shown. NdNiO$_2$ displays clear differences in its Fermiology from CaCuO$_2$.  The Nd $d_{3z^2-r^2}$ orbital gives rise to a band that crosses the chemical potential, leading to an oblate Fermi surface centered on the $\Gamma$ point (see panels for $k_z = 0.0$ and $0.125$), while the  Nd $d_{xy}$ orbital gives rise to an oblate Fermi surface pocket centered on $A$ (see panels for $k_z = 0.375$, $0.45$, and $0.5$), as also found in previous DFT and DFT+U calculations~\cite{Lee04,Liu14,botana2019} and in a recent DFT+DMFT study~\cite{lechermann2019late}.  Interestingly, in DFT calculations for  LaNiO$_2$ the pocket at $\Gamma$ is noticeably smaller than in NdNiO$_2$~\cite{Lee04} and it is absent in DFT+DMFT calculations for the La material~\cite{si2019topotactic}.

The Nd-derived bands centered at $\Gamma$ and $A$ cross the Fermi level and contain electrons, thereby doping the Ni-$d$-derived band. The amount of doping may be obtained from the areas of the Fermi surfaces; we find that the two Nd-derived bands each contain about $0.05$ electrons (summed over spin) yielding a total doping of the Ni-derived band of about $0.1$ holes. This doping is substantially smaller than the $0.5$ total charge transfer to the Nd found in the wide energy window calculation, and only about half of the electron transfer from the Ni $d_{x^2-y^2}$ orbitals respectively. This doping is also larger than what was found for LaNiO$_2$ in DFT calculations~\cite{botana2019}; the difference is a result of the different electronegativities of La and Nd. 

\begin{figure}[t]
\centering
\includegraphics[width=0.48\columnwidth]{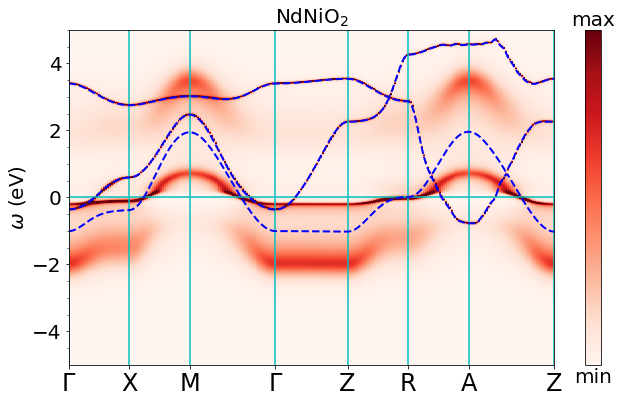}
\includegraphics[width=0.48\columnwidth]{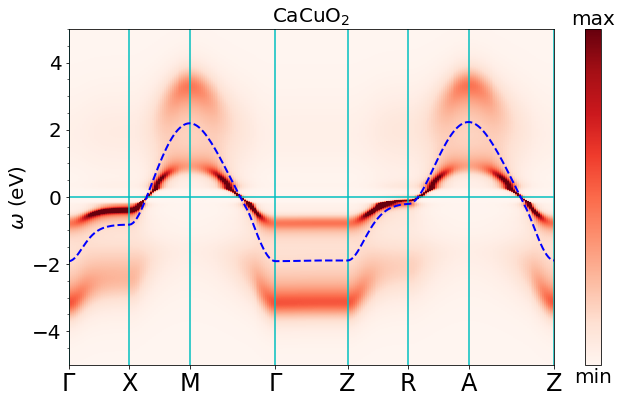}
\caption{Frequency dependence of the spectral function $A(k,\omega)$ plotted along indicated lines in momentum space  (false color), computed from narrow energy window paramagnetic phase DMFT with $U$ = \SI{3.1}{eV} at temperature $T = \SI{290}{K}$, along with uncorrelated Wannier bands (dashed blue lines) for NdNiO$_2$ (left panel) and CaCuO$_2$ (right panel).  }
\label{Bands}
\end{figure}

The wide energy window calculations show that while these Nd derived bands have some admixture of Ni-d character, the band basis self-energies are quite small ($Z > 0.9$ in the vicinity of the Fermi level). Further, analysis of the Wannier Hamiltonian in the narrow energy window shows that the hybridization between these bands and the Ni $d_{x^2-y^2}$ derived band is small (see Tab.~\ref{tab:narrow_params}), and thus the bands are effectively decoupled (as may also be seen from the lack of level repulsion where the bands cross). To further investigate the effect of hybridization with the Nd bands, we perform a one-band Wannier fit capturing only the Ni $d_{x^2-y^2}$-derived band (thus freezing out any dynamical charge transfer) and recompute the self-energy with the occupancy of the band set equal to the self-doped value.  No significant differences are found (a comparison of the two calculations is shown in the Supplementary Material). This key insight proves that the low-energy physics of the nickelate material can be understood as a (doped) single-band Hubbard model,  similar to the physics of cuprates. The important difference is, however, the ``self-doping'' effect of the Nd-derived bands which mainly act as an electron reservoir.

Additionally, the self-doping of the Ni-derived orbital makes it acquire a non-negligible $k_z$ dispersion, leading to a region of momentum space where the Ni $d_{x^2-y^2}$-derived Fermi surface passes through the van Hove point (see Fig.~\ref{fermi_surface}). The resulting enhanced density of states leads to more pronounced correlation effects in the Ni material; for example, it is clearly seen that the Fermi surface is substantially broadened when the van Hove point is near the Fermi level.

In Fig.~\ref{Bands}, we compare the DFT bands (blue dashed lines) to the momentum-resolved correlated spectral function (false color) computed at $U$ = \SI{3.1}{eV} in the paramagnetic phase for both materials. The results for CaCuO$_2$ show the expected behavior: a quasiparticle band renormalized by about a factor of 2 with respect to the DFT bands, and a shake-off feature, below the band for occupied states and above for empty states. The results for NdNiO$_2$ show similar shake-off features and renormalization structure (with a somewhat larger renormalization constant). An important difference is that even at $k_z=0$ the van Hove singularity near the $X$ point is much closer to the Fermi level in the Ni material than in the Cu one; this difference is in essence a doping effect, enhanced by the stronger correlations and the $k_z$ dispersion. The renormalization is strong enough that the bottom of the Nd band at $A$ lies substantially ($\sim \SI{1}{eV}$) below the bottom of the quasiparticle band, and even the bottom of the shallower Nd band at $\Gamma$ is at about the same energy as the bottom of the quasiparticle band. This suggests that it might be difficult to dope NdNiO$_2$ to the point where the Nd states are fully emptied out. 

\begin{figure*}[t]
\centering
\includegraphics[width = \textwidth]{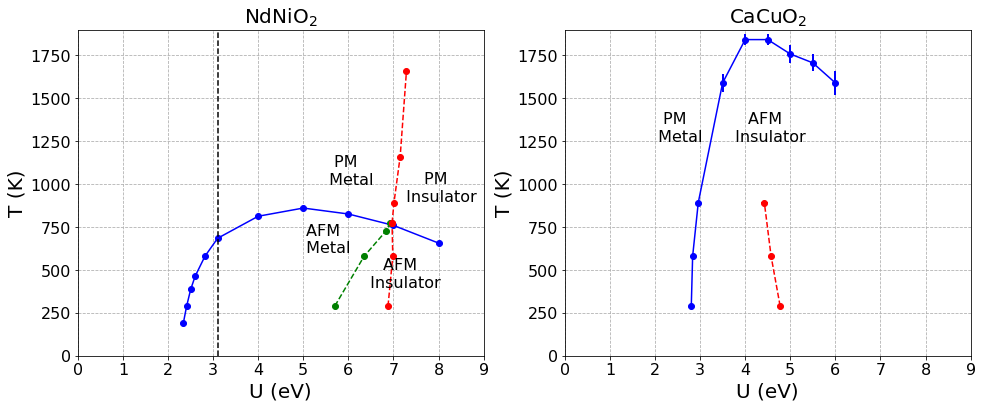}
\caption{Phase diagrams of NdNiO$_2$ (left panel) and CaCuO$_2$ (right panel) in the plane of temperature $T$ and on-site interaction $U$, calculated using single-site DMFT with band parameters obtained from a narrow energy window Wannier construction.  The solid blue line shows the paramagnetic-antiferromagnetic transition. The red dashed lines show the paramagnetic metal to paramagnetic insulator transition, and in the left panel the green dashed line shows the antiferromagnetic metal to antiferromagnetic insulator transition. In the antiferromagnetic regime (temperatures below the blue line) the red dashed line indicates the paramagnetic metal-insulator transition if the antiferromagnetic solution is suppressed. Note that the end points do not correspond to critical points, but show the parameter range in which we have performed calculations. The dashed black vertical line marks the approximate narrow-window interaction parameter found in cRPA calculations~\cite{nomura2019}.}
\label{phase_diagram}
\end{figure*}

We now explore a wider range of interactions in the frontier orbital model, presenting in Fig.~\ref{phase_diagram} the metal-insulator and paramagnetic-antiferromagnetic phase transitions calculated for the two materials for a range of interaction strengths. The obtained magnetic transition temperatures are high, consistent with the known DMFT overestimation of the magnetic transition temperature by a factor of 2-3~\cite{Neel01,Neel02,Neel03,Neel04}. This overestimation is especially large in quasi two-dimensional systems where spatial fluctuations can play an important role. Still, the calculated transition temperatures set a scale at which magnetic correlations become relevant.

The calculated cuprate phase diagram is consistent with the large body of existing literature on the one-band Hubbard model~\cite{Neel01}:  The transition temperature is high enough that the theoretically predicted paramagnetic metal-paramagnetic insulator transition is preempted by the antiferromagnetic phase. The calculated N{\'e}el temperature is maximized at $U \approx$ \SI{4}{eV} $\approx$ the bandwidth; this scale is a demarcation between Mott insulating physics and intermediate correlations; the wide energy window calculations suggest that the cuprate materials are on the moderately correlated side of this crossover.  

The nickelate phase diagram differs in several aspects. The  critical $U$ required for the onset of magnetism is smaller than in the cuprate case, corresponding to the smaller bandwidth of the Ni compound; however, the ratio of the critical $U$ to the bandwidth is about the same in the two cases. In the large $U$ regime of the phase diagram, where the model is insulating at all temperatures, the N{\'e}el temperature of the Ni material is lower than that of the cuprate by about a factor of two, consistent with the $3:4$ ratio of nearest neighbor hoppings and the superexchange estimate $J=t^2/U$.

The most important difference between the two phase diagrams, however, is that the self-doping provided by the Nd $5d$ bands means that up to a moderately large $U$, the Ni material is an antiferromagnetic metal, as found in other DMFT calculations~\cite{gu2019}. Only if $U$ is increased beyond a large, temperature-dependent value of about $\SI{5}{eV}$ is an antiferromagnetic insulator phase found in this calculation. The physics of the transition is that the antiferromagnetic order parameter opens a gap in the Ni spectral function; when this gap becomes large enough so that the top of the lower antiferromagnetic band falls below the bottom of the Nd $5d$ bands, then all of the charge from the Nd $5d$ states is transferred to the Ni band, which becomes half-filled such that an insulating state may result. The transition in the paramagnetic phase is similarly driven when the splitting between the upper and lower Hubbard band becomes large enough. This transition depends crucially on the energetics of Ni-Nd charge transfer, which, however, may not be correctly captured by the three-band frontier orbital model used here.  It may also be preempted by a high-spin/low-spin transition in which holes are added to the Ni-$d_{3z^2-r^2}$ sector.  Both of these effects will tend to push the transition to larger $U$. Understanding these issues is a theoretically interesting question requiring a fully charge self-consistent calculation and a theoretically justified double counting correction.   

{\bf Conclusion} We have presented a comparative study of two isostructural materials, NdNiO$_2$ and CaCuO$_2$. Both have a transition metal configuration near $d^9$ and correlation physics dominated by the transition metal $d_{x^2-y^2}$ orbital, and both should be regarded as charge transfer materials with a significant admixture of an oxygen hole/$d^{10}$ configuration and a negligible role of the $d^{8}$ configuration. A salient difference between the two materials is that the energy splitting between the oxygen $2p$ and transition metal $3d$ bands is less in the cuprate than in the nickelate material, leading to relatively weaker correlations in the cuprate case. Previous work of Weber and collaborators \cite{Weber2012} on cuprates suggests that the charge transfer gap is anticorrelated with the superconducting transition temperature, suggesting that the nickelate material will have a lower superconducting transition temperature than its cuprate analogue \cite{botana2019}. However, this difference appears to be quantitative and not a fundamental difference between the physics of these two materials. This conclusion is different from that found in DFT+U calculations at $U\gtrsim \SI{6}{eV}$~\cite{Liu14, botana2019} and DFT+DMFT calculations at $U\gtrsim \SI{10}{eV}$~\cite{lechermann2019late}.

Other differences between the materials are a modest decrease in the proportion of oxygen holes and, most importantly, that in the nickelate material bands of Nd $5d$ origin cross the Fermi level, so that charge is transferred from the Ni-$d$/O-$p$ complex to the Nd bands, leading to a ``self-doped'' situation. The Nd states also hybridize with the Ni/O states, leading to a non-negligible $k_z$ dispersion and the presence of a van Hove singularity that crosses the chemical potential as $k_z$ is varied. We find however that dynamical charge transfer to Nd is not important: one obtains the same self-energy in a one-band model with the same Fermi surface for the Ni-derived band. Among other things, this means that the interesting Kondo effects discussed in Ref.~\onlinecite{zhang2019kondo} are not relevant to the present calculation.

Perhaps the greatest theoretical uncertainties in the calculated results arise from the value of $U$ and from the double counting correction. The double counting correction, which controls the placement in energy of the correlated states with respect to the uncorrelated ones, is of particular importance. In the cuprate case the standard double counting  leads to a  higher overlap between the O and Cu states, and thus to weaker correlations, while adjusting the double counting leads to weaker Cu-O overlap, increasing the correlations in the frontier orbitals. In the Ni case adjusting the double counting would also affect the charge transfer to the Nd states. The $U$ value affects whether the $d_{x^2-y^2}$ orbitals can be driven into a Mott state~\cite{lechermann2019late}. Despite these uncertainties, it is clear that the Ni material is more strongly correlated than the Cu material. Other uncertainties, including whether to include the other (almost completely filled) Ni $d$-orbitals, Nd-$d$ orbitals, or O-$p$ orbitals as correlated states, seem to us to have much less effect.  

Our conclusions are at variance with others reported in the literature. Several authors argue for the importance of the high-spin $d^8$ state~\cite{Liu14,botana2019,lechermann2019late}, which combined with an orbital selective Mott transition leads to a picture of a Nd/Ni hybridized band Kondo-hybridized to local moments on the Ni sites. Our belief is that this physics is unlikely to be relevant to the 112 nickelates. However, higher-$U$ parameter regimes the models we and others have derived for the Ni materials display very interesting physics, including the Kondo and multi-orbital effects discussed in Ref.~\cite{zhang2019selfdoped,werner2019nickelate} and the high-spin/low-spin and orbitally selective Mott physics found in DFT+U~\cite{Liu14,botana2019} and DFT+DMFT~\cite{lechermann2019late}, which warrant further investigation.

Most importantly, we find that the low-energy physics of the Ni material is described by a three-band model, with one correlated Ni-$d_{x^2-y^2}$-derived band with a van Hove singularity that crosses the Fermi level as $k_z$ is varied and two Nd-derived weakly correlated spectator bands. We argue that the main effect of these spectator bands is a self-doping of the Ni-$d_{x^2-y^2}$-derived band. Thus, a fundamental insight of our work is that the low-energy physics of NdNiO$_2$ can be modeled by a doped single-band Hubbard model, which is (besides the doping) similar to the one for CaCuO$_2$, albeit with a larger charge transfer gap and thus likely weaker superconductivity upon Sr doping \cite{Weber2012}.

The effect of the van Hove singularity and the spectator bands on the superconducting physics of the Ni-derived band remains to be determined. For example, one may ask, within the theory as defined here, why the (self-) doped Ni compound is not superconducting even in the nominally stoichiometric case~\cite{Hwang2019}. Cluster dynamical mean-field calculations to answer this question are in progress. The evolution of quasiparticle properties and long-range order as a function of doping is also an important open question.

\section{Acknowledgements} We thank A.~Georges and F. Lechermann for very helpful discussions. JK thanks A.~B.~Georgescu and H.~U.~R.~Strand for assistance with DFT and DMFT calculations. This work was supported by the Materials Sciences and Engineering Division, Basic Energy Sciences, Office of Science, US DOE. ASB thanks ASU for startup funds. The Flatiron Institute is a division of the Simons Foundation.
\bibliography{references.bib}

\begin{thebibliography}{70}%
\makeatletter
\providecommand \@ifxundefined [1]{%
 \@ifx{#1\undefined}
}%
\providecommand \@ifnum [1]{%
 \ifnum #1\expandafter \@firstoftwo
 \else \expandafter \@secondoftwo
 \fi
}%
\providecommand \@ifx [1]{%
 \ifx #1\expandafter \@firstoftwo
 \else \expandafter \@secondoftwo
 \fi
}%
\providecommand \natexlab [1]{#1}%
\providecommand \enquote  [1]{``#1''}%
\providecommand \bibnamefont  [1]{#1}%
\providecommand \bibfnamefont [1]{#1}%
\providecommand \citenamefont [1]{#1}%
\providecommand \href@noop [0]{\@secondoftwo}%
\providecommand \href [0]{\begingroup \@sanitize@url \@href}%
\providecommand \@href[1]{\@@startlink{#1}\@@href}%
\providecommand \@@href[1]{\endgroup#1\@@endlink}%
\providecommand \@sanitize@url [0]{\catcode `\\12\catcode `\$12\catcode
  `\&12\catcode `\#12\catcode `\^12\catcode `\_12\catcode `\%12\relax}%
\providecommand \@@startlink[1]{}%
\providecommand \@@endlink[0]{}%
\providecommand \url  [0]{\begingroup\@sanitize@url \@url }%
\providecommand \@url [1]{\endgroup\@href {#1}{\urlprefix }}%
\providecommand \urlprefix  [0]{URL }%
\providecommand \Eprint [0]{\href }%
\providecommand \doibase [0]{http://dx.doi.org/}%
\providecommand \selectlanguage [0]{\@gobble}%
\providecommand \bibinfo  [0]{\@secondoftwo}%
\providecommand \bibfield  [0]{\@secondoftwo}%
\providecommand \translation [1]{[#1]}%
\providecommand \BibitemOpen [0]{}%
\providecommand \bibitemStop [0]{}%
\providecommand \bibitemNoStop [0]{.\EOS\space}%
\providecommand \EOS [0]{\spacefactor3000\relax}%
\providecommand \BibitemShut  [1]{\csname bibitem#1\endcsname}%
\let\auto@bib@innerbib\@empty
\bibitem [{\citenamefont {Anderson}(1987)}]{Anderson87}%
  \BibitemOpen
  \bibfield  {author} {\bibinfo {author} {\bibfnamefont {P.~W.}\ \bibnamefont
  {Anderson}},\ }\href {\doibase 10.1126/science.235.4793.1196} {\bibfield
  {journal} {\bibinfo  {journal} {Science}\ }\textbf {\bibinfo {volume}
  {235}},\ \bibinfo {pages} {1196} (\bibinfo {year} {1987})}\BibitemShut
  {NoStop}%
\bibitem [{\citenamefont {Pickett}(1989)}]{Pickett89}%
  \BibitemOpen
  \bibfield  {author} {\bibinfo {author} {\bibfnamefont {W.~E.}\ \bibnamefont
  {Pickett}},\ }\href {\doibase 10.1103/RevModPhys.61.433} {\bibfield
  {journal} {\bibinfo  {journal} {Rev. Mod. Phys.}\ }\textbf {\bibinfo {volume}
  {61}},\ \bibinfo {pages} {433} (\bibinfo {year} {1989})}\BibitemShut
  {NoStop}%
\bibitem [{\citenamefont {Keimer}\ \emph {et~al.}(2015)\citenamefont {Keimer},
  \citenamefont {Kivelson}, \citenamefont {Norman}, \citenamefont {Uchida},\
  and\ \citenamefont {Zaanen}}]{Keimer15}%
  \BibitemOpen
  \bibfield  {author} {\bibinfo {author} {\bibfnamefont {B.}~\bibnamefont
  {Keimer}}, \bibinfo {author} {\bibfnamefont {S.~A.}\ \bibnamefont
  {Kivelson}}, \bibinfo {author} {\bibfnamefont {M.~R.}\ \bibnamefont
  {Norman}}, \bibinfo {author} {\bibfnamefont {S.}~\bibnamefont {Uchida}}, \
  and\ \bibinfo {author} {\bibfnamefont {J.}~\bibnamefont {Zaanen}},\ }\href
  {\doibase 10.1038/nature14165} {\bibfield  {journal} {\bibinfo  {journal}
  {Nature}\ }\textbf {\bibinfo {volume} {518}},\ \bibinfo {pages} {179}
  (\bibinfo {year} {2015})}\BibitemShut {NoStop}%
\bibitem [{\citenamefont {Proust}\ and\ \citenamefont
  {Taillefer}(2019)}]{Proust19}%
  \BibitemOpen
  \bibfield  {author} {\bibinfo {author} {\bibfnamefont {C.}~\bibnamefont
  {Proust}}\ and\ \bibinfo {author} {\bibfnamefont {L.}~\bibnamefont
  {Taillefer}},\ }\href {\doibase 10.1146/annurev-conmatphys-031218-013210}
  {\bibfield  {journal} {\bibinfo  {journal} {Annu. Rev. Condens. Matter
  Phys.}\ }\textbf {\bibinfo {volume} {10}},\ \bibinfo {pages} {409} (\bibinfo
  {year} {2019})}\BibitemShut {NoStop}%
\bibitem [{\citenamefont {Bednorz}\ and\ \citenamefont
  {M{\"u}ller}(1986)}]{Bednorz86}%
  \BibitemOpen
  \bibfield  {author} {\bibinfo {author} {\bibfnamefont {J.~G.}\ \bibnamefont
  {Bednorz}}\ and\ \bibinfo {author} {\bibfnamefont {K.~A.}\ \bibnamefont
  {M{\"u}ller}},\ }\href {\doibase https://doi.org/10.1007/BF01303701}
  {\bibfield  {journal} {\bibinfo  {journal} {Z. Phys. B Condensed Matter}\
  }\textbf {\bibinfo {volume} {64}},\ \bibinfo {pages} {189} (\bibinfo {year}
  {1986})}\BibitemShut {NoStop}%
\bibitem [{\citenamefont {Emery}(1987)}]{Emery87}%
  \BibitemOpen
  \bibfield  {author} {\bibinfo {author} {\bibfnamefont {V.~J.}\ \bibnamefont
  {Emery}},\ }\href {\doibase 10.1103/PhysRevLett.58.2794} {\bibfield
  {journal} {\bibinfo  {journal} {Phys. Rev. Lett.}\ }\textbf {\bibinfo
  {volume} {58}},\ \bibinfo {pages} {2794} (\bibinfo {year}
  {1987})}\BibitemShut {NoStop}%
\bibitem [{\citenamefont {Littlewood}\ \emph {et~al.}(1989)\citenamefont
  {Littlewood}, \citenamefont {Varma},\ and\ \citenamefont
  {Abrahams}}]{Littlewood89}%
  \BibitemOpen
  \bibfield  {author} {\bibinfo {author} {\bibfnamefont {P.~B.}\ \bibnamefont
  {Littlewood}}, \bibinfo {author} {\bibfnamefont {C.~M.}\ \bibnamefont
  {Varma}}, \ and\ \bibinfo {author} {\bibfnamefont {E.}~\bibnamefont
  {Abrahams}},\ }\href {\doibase 10.1103/PhysRevLett.63.2602} {\bibfield
  {journal} {\bibinfo  {journal} {Phys. Rev. Lett.}\ }\textbf {\bibinfo
  {volume} {63}},\ \bibinfo {pages} {2602} (\bibinfo {year}
  {1989})}\BibitemShut {NoStop}%
\bibitem [{\citenamefont {Anisimov}\ \emph {et~al.}(1999)\citenamefont
  {Anisimov}, \citenamefont {Bukhvalov},\ and\ \citenamefont
  {Rice}}]{Anisimov99}%
  \BibitemOpen
  \bibfield  {author} {\bibinfo {author} {\bibfnamefont {V.~I.}\ \bibnamefont
  {Anisimov}}, \bibinfo {author} {\bibfnamefont {D.}~\bibnamefont {Bukhvalov}},
  \ and\ \bibinfo {author} {\bibfnamefont {T.~M.}\ \bibnamefont {Rice}},\
  }\href {\doibase 10.1103/PhysRevB.59.7901} {\bibfield  {journal} {\bibinfo
  {journal} {Phys. Rev. B}\ }\textbf {\bibinfo {volume} {59}},\ \bibinfo
  {pages} {7901} (\bibinfo {year} {1999})}\BibitemShut {NoStop}%
\bibitem [{\citenamefont {Lee}\ and\ \citenamefont {Pickett}(2004)}]{Lee04}%
  \BibitemOpen
  \bibfield  {author} {\bibinfo {author} {\bibfnamefont {K.-W.}\ \bibnamefont
  {Lee}}\ and\ \bibinfo {author} {\bibfnamefont {W.~E.}\ \bibnamefont
  {Pickett}},\ }\href {\doibase 10.1103/PhysRevB.70.165109} {\bibfield
  {journal} {\bibinfo  {journal} {Phys. Rev. B}\ }\textbf {\bibinfo {volume}
  {70}},\ \bibinfo {pages} {165109} (\bibinfo {year} {2004})}\BibitemShut
  {NoStop}%
\bibitem [{\citenamefont {Liu}\ \emph {et~al.}(2014)\citenamefont {Liu},
  \citenamefont {Wu}, \citenamefont {Jia}, \citenamefont {Zhang}, \citenamefont
  {Zeng}, \citenamefont {Lin},\ and\ \citenamefont {Li}}]{Liu14}%
  \BibitemOpen
  \bibfield  {author} {\bibinfo {author} {\bibfnamefont {T.}~\bibnamefont
  {Liu}}, \bibinfo {author} {\bibfnamefont {H.}~\bibnamefont {Wu}}, \bibinfo
  {author} {\bibfnamefont {T.}~\bibnamefont {Jia}}, \bibinfo {author}
  {\bibfnamefont {X.}~\bibnamefont {Zhang}}, \bibinfo {author} {\bibfnamefont
  {Z.}~\bibnamefont {Zeng}}, \bibinfo {author} {\bibfnamefont {H.~Q.}\
  \bibnamefont {Lin}}, \ and\ \bibinfo {author} {\bibfnamefont {X.~G.}\
  \bibnamefont {Li}},\ }\href {\doibase 10.1063/1.4873537} {\bibfield
  {journal} {\bibinfo  {journal} {AIP Advances}\ }\textbf {\bibinfo {volume}
  {4}},\ \bibinfo {pages} {047132} (\bibinfo {year} {2014})}\BibitemShut
  {NoStop}%
\bibitem [{\citenamefont {Botana}\ and\ \citenamefont
  {Norman}(2018)}]{Botana18}%
  \BibitemOpen
  \bibfield  {author} {\bibinfo {author} {\bibfnamefont {A.~S.}\ \bibnamefont
  {Botana}}\ and\ \bibinfo {author} {\bibfnamefont {M.~R.}\ \bibnamefont
  {Norman}},\ }\href {\doibase 10.1103/PhysRevMaterials.2.104803} {\bibfield
  {journal} {\bibinfo  {journal} {Phys. Rev. Mat.}\ }\textbf {\bibinfo {volume}
  {2}},\ \bibinfo {pages} {104803} (\bibinfo {year} {2018})}\BibitemShut
  {NoStop}%
\bibitem [{\citenamefont {Chaloupka}\ and\ \citenamefont
  {Khaliullin}(2008)}]{Chaloupka08}%
  \BibitemOpen
  \bibfield  {author} {\bibinfo {author} {\bibfnamefont {J.}~\bibnamefont
  {Chaloupka}}\ and\ \bibinfo {author} {\bibfnamefont {G.}~\bibnamefont
  {Khaliullin}},\ }\href {\doibase 10.1103/PhysRevLett.100.016404} {\bibfield
  {journal} {\bibinfo  {journal} {Phys. Rev. Lett.}\ }\textbf {\bibinfo
  {volume} {100}},\ \bibinfo {pages} {016404} (\bibinfo {year}
  {2008})}\BibitemShut {NoStop}%
\bibitem [{\citenamefont {Zhang}\ \emph {et~al.}(2017)\citenamefont {Zhang},
  \citenamefont {Botana}, \citenamefont {Freeland}, \citenamefont {Phelan},
  \citenamefont {Zheng}, \citenamefont {Pardo}, \citenamefont {Norman},\ and\
  \citenamefont {Mitchell}}]{Zhang2017}%
  \BibitemOpen
  \bibfield  {author} {\bibinfo {author} {\bibfnamefont {J.}~\bibnamefont
  {Zhang}}, \bibinfo {author} {\bibfnamefont {A.~S.}\ \bibnamefont {Botana}},
  \bibinfo {author} {\bibfnamefont {J.~W.}\ \bibnamefont {Freeland}}, \bibinfo
  {author} {\bibfnamefont {D.}~\bibnamefont {Phelan}}, \bibinfo {author}
  {\bibfnamefont {H.}~\bibnamefont {Zheng}}, \bibinfo {author} {\bibfnamefont
  {V.}~\bibnamefont {Pardo}}, \bibinfo {author} {\bibfnamefont {M.~R.}\
  \bibnamefont {Norman}}, \ and\ \bibinfo {author} {\bibfnamefont {J.~F.}\
  \bibnamefont {Mitchell}},\ }\href {https://doi.org/10.1038/nphys4149}
  {\bibfield  {journal} {\bibinfo  {journal} {Nat. Phys.}\ }\textbf {\bibinfo
  {volume} {13}},\ \bibinfo {pages} {864} (\bibinfo {year} {2017})}\BibitemShut
  {NoStop}%
\bibitem [{\citenamefont {Li}\ \emph {et~al.}(2019)\citenamefont {Li},
  \citenamefont {Lee}, \citenamefont {Wang}, \citenamefont {Osada},
  \citenamefont {Crossley}, \citenamefont {Lee}, \citenamefont {Cui},
  \citenamefont {Hikita},\ and\ \citenamefont {Hwang}}]{Hwang2019}%
  \BibitemOpen
  \bibfield  {author} {\bibinfo {author} {\bibfnamefont {D.}~\bibnamefont
  {Li}}, \bibinfo {author} {\bibfnamefont {K.}~\bibnamefont {Lee}}, \bibinfo
  {author} {\bibfnamefont {B.~Y.}\ \bibnamefont {Wang}}, \bibinfo {author}
  {\bibfnamefont {M.}~\bibnamefont {Osada}}, \bibinfo {author} {\bibfnamefont
  {S.}~\bibnamefont {Crossley}}, \bibinfo {author} {\bibfnamefont {H.~R.}\
  \bibnamefont {Lee}}, \bibinfo {author} {\bibfnamefont {Y.}~\bibnamefont
  {Cui}}, \bibinfo {author} {\bibfnamefont {Y.}~\bibnamefont {Hikita}}, \ and\
  \bibinfo {author} {\bibfnamefont {H.~Y.}\ \bibnamefont {Hwang}},\ }\href
  {\doibase 10.1038/s41586-019-1496-5} {\bibfield  {journal} {\bibinfo
  {journal} {Nature}\ }\textbf {\bibinfo {volume} {572}},\ \bibinfo {pages}
  {624} (\bibinfo {year} {2019})}\BibitemShut {NoStop}%
\bibitem [{\citenamefont {Botana}\ and\ \citenamefont
  {Norman}(2020)}]{botana2019}%
  \BibitemOpen
  \bibfield  {author} {\bibinfo {author} {\bibfnamefont {A.~S.}\ \bibnamefont
  {Botana}}\ and\ \bibinfo {author} {\bibfnamefont {M.~R.}\ \bibnamefont
  {Norman}},\ }\href {\doibase 10.1103/PhysRevX.10.011024} {\bibfield
  {journal} {\bibinfo  {journal} {Phys. Rev. X}\ }\textbf {\bibinfo {volume}
  {10}},\ \bibinfo {pages} {011024} (\bibinfo {year} {2020})}\BibitemShut
  {NoStop}%
\bibitem [{\citenamefont {Nomura}\ \emph {et~al.}(2019)\citenamefont {Nomura},
  \citenamefont {Hirayama}, \citenamefont {Tadano}, \citenamefont {Yoshimoto},
  \citenamefont {Nakamura},\ and\ \citenamefont {Arita}}]{nomura2019}%
  \BibitemOpen
  \bibfield  {author} {\bibinfo {author} {\bibfnamefont {Y.}~\bibnamefont
  {Nomura}}, \bibinfo {author} {\bibfnamefont {M.}~\bibnamefont {Hirayama}},
  \bibinfo {author} {\bibfnamefont {T.}~\bibnamefont {Tadano}}, \bibinfo
  {author} {\bibfnamefont {Y.}~\bibnamefont {Yoshimoto}}, \bibinfo {author}
  {\bibfnamefont {K.}~\bibnamefont {Nakamura}}, \ and\ \bibinfo {author}
  {\bibfnamefont {R.}~\bibnamefont {Arita}},\ }\href {\doibase
  10.1103/PhysRevB.100.205138} {\bibfield  {journal} {\bibinfo  {journal}
  {Phys. Rev. B}\ }\textbf {\bibinfo {volume} {100}},\ \bibinfo {pages}
  {205138} (\bibinfo {year} {2019})}\BibitemShut {NoStop}%
\bibitem [{\citenamefont {Hepting}\ \emph {et~al.}(2020)\citenamefont
  {Hepting}, \citenamefont {Li}, \citenamefont {Jia}, \citenamefont {Lu},
  \citenamefont {Paris}, \citenamefont {Tseng}, \citenamefont {Feng},
  \citenamefont {Osada}, \citenamefont {Been}, \citenamefont {Hikita},
  \citenamefont {Chuang}, \citenamefont {Hussain}, \citenamefont {Zhou},
  \citenamefont {Nag}, \citenamefont {Garcia-Fernandez}, \citenamefont {Rossi},
  \citenamefont {Huang}, \citenamefont {Huang}, \citenamefont {Shen},
  \citenamefont {Schmitt}, \citenamefont {Hwang}, \citenamefont {Moritz},
  \citenamefont {Zaanen}, \citenamefont {Devereaux},\ and\ \citenamefont
  {Lee}}]{hepting2019}%
  \BibitemOpen
  \bibfield  {author} {\bibinfo {author} {\bibfnamefont {M.}~\bibnamefont
  {Hepting}}, \bibinfo {author} {\bibfnamefont {D.}~\bibnamefont {Li}},
  \bibinfo {author} {\bibfnamefont {C.~J.}\ \bibnamefont {Jia}}, \bibinfo
  {author} {\bibfnamefont {H.}~\bibnamefont {Lu}}, \bibinfo {author}
  {\bibfnamefont {E.}~\bibnamefont {Paris}}, \bibinfo {author} {\bibfnamefont
  {Y.}~\bibnamefont {Tseng}}, \bibinfo {author} {\bibfnamefont
  {X.}~\bibnamefont {Feng}}, \bibinfo {author} {\bibfnamefont {M.}~\bibnamefont
  {Osada}}, \bibinfo {author} {\bibfnamefont {E.}~\bibnamefont {Been}},
  \bibinfo {author} {\bibfnamefont {Y.}~\bibnamefont {Hikita}}, \bibinfo
  {author} {\bibfnamefont {Y.~D.}\ \bibnamefont {Chuang}}, \bibinfo {author}
  {\bibfnamefont {Z.}~\bibnamefont {Hussain}}, \bibinfo {author} {\bibfnamefont
  {K.~J.}\ \bibnamefont {Zhou}}, \bibinfo {author} {\bibfnamefont
  {A.}~\bibnamefont {Nag}}, \bibinfo {author} {\bibfnamefont {M.}~\bibnamefont
  {Garcia-Fernandez}}, \bibinfo {author} {\bibfnamefont {M.}~\bibnamefont
  {Rossi}}, \bibinfo {author} {\bibfnamefont {H.~Y.}\ \bibnamefont {Huang}},
  \bibinfo {author} {\bibfnamefont {D.~J.}\ \bibnamefont {Huang}}, \bibinfo
  {author} {\bibfnamefont {Z.~X.}\ \bibnamefont {Shen}}, \bibinfo {author}
  {\bibfnamefont {T.}~\bibnamefont {Schmitt}}, \bibinfo {author} {\bibfnamefont
  {H.~Y.}\ \bibnamefont {Hwang}}, \bibinfo {author} {\bibfnamefont
  {B.}~\bibnamefont {Moritz}}, \bibinfo {author} {\bibfnamefont
  {J.}~\bibnamefont {Zaanen}}, \bibinfo {author} {\bibfnamefont {T.~P.}\
  \bibnamefont {Devereaux}}, \ and\ \bibinfo {author} {\bibfnamefont {W.~S.}\
  \bibnamefont {Lee}},\ }\href {\doibase 10.1038/s41563-019-0585-z} {\bibfield
  {journal} {\bibinfo  {journal} {Nat. Mater.}\ } (\bibinfo {year} {2020}),\
  10.1038/s41563-019-0585-z}\BibitemShut {NoStop}%
\bibitem [{\citenamefont {Wu}\ \emph {et~al.}(2020)\citenamefont {Wu},
  \citenamefont {Di~Sante}, \citenamefont {Schwemmer}, \citenamefont {Hanke},
  \citenamefont {Hwang}, \citenamefont {Raghu},\ and\ \citenamefont
  {Thomale}}]{wu2019}%
  \BibitemOpen
  \bibfield  {author} {\bibinfo {author} {\bibfnamefont {X.}~\bibnamefont
  {Wu}}, \bibinfo {author} {\bibfnamefont {D.}~\bibnamefont {Di~Sante}},
  \bibinfo {author} {\bibfnamefont {T.}~\bibnamefont {Schwemmer}}, \bibinfo
  {author} {\bibfnamefont {W.}~\bibnamefont {Hanke}}, \bibinfo {author}
  {\bibfnamefont {H.~Y.}\ \bibnamefont {Hwang}}, \bibinfo {author}
  {\bibfnamefont {S.}~\bibnamefont {Raghu}}, \ and\ \bibinfo {author}
  {\bibfnamefont {R.}~\bibnamefont {Thomale}},\ }\href {\doibase
  10.1103/PhysRevB.101.060504} {\bibfield  {journal} {\bibinfo  {journal}
  {Phys. Rev. B}\ }\textbf {\bibinfo {volume} {101}},\ \bibinfo {pages}
  {060504} (\bibinfo {year} {2020})}\BibitemShut {NoStop}%
\bibitem [{\citenamefont {Sakakibara}\ \emph {et~al.}()\citenamefont
  {Sakakibara}, \citenamefont {Usui}, \citenamefont {Suzuki}, \citenamefont
  {Kotani}, \citenamefont {Aoki},\ and\ \citenamefont
  {Kuroki}}]{sakakibara2019}%
  \BibitemOpen
  \bibfield  {author} {\bibinfo {author} {\bibfnamefont {H.}~\bibnamefont
  {Sakakibara}}, \bibinfo {author} {\bibfnamefont {H.}~\bibnamefont {Usui}},
  \bibinfo {author} {\bibfnamefont {K.}~\bibnamefont {Suzuki}}, \bibinfo
  {author} {\bibfnamefont {T.}~\bibnamefont {Kotani}}, \bibinfo {author}
  {\bibfnamefont {H.}~\bibnamefont {Aoki}}, \ and\ \bibinfo {author}
  {\bibfnamefont {K.}~\bibnamefont {Kuroki}},\ }\href@noop {} {\ }\Eprint
  {http://arxiv.org/abs/1909.00060} {arXiv:1909.00060} \BibitemShut {NoStop}%
\bibitem [{\citenamefont {Gao}\ \emph {et~al.}()\citenamefont {Gao},
  \citenamefont {Wang}, \citenamefont {Fang},\ and\ \citenamefont
  {Weng}}]{gao2019electronic}%
  \BibitemOpen
  \bibfield  {author} {\bibinfo {author} {\bibfnamefont {J.}~\bibnamefont
  {Gao}}, \bibinfo {author} {\bibfnamefont {Z.}~\bibnamefont {Wang}}, \bibinfo
  {author} {\bibfnamefont {C.}~\bibnamefont {Fang}}, \ and\ \bibinfo {author}
  {\bibfnamefont {H.}~\bibnamefont {Weng}},\ }\href@noop {} {\ }\Eprint
  {http://arxiv.org/abs/1909.04657} {arXiv:1909.04657} \BibitemShut {NoStop}%
\bibitem [{\citenamefont {Zhang}\ \emph
  {et~al.}(2020{\natexlab{a}})\citenamefont {Zhang}, \citenamefont {Jin},
  \citenamefont {Wang}, \citenamefont {Xi}, \citenamefont {Shi}, \citenamefont
  {Ye},\ and\ \citenamefont {Mei}}]{zhang2019effective}%
  \BibitemOpen
  \bibfield  {author} {\bibinfo {author} {\bibfnamefont {H.}~\bibnamefont
  {Zhang}}, \bibinfo {author} {\bibfnamefont {L.}~\bibnamefont {Jin}}, \bibinfo
  {author} {\bibfnamefont {S.}~\bibnamefont {Wang}}, \bibinfo {author}
  {\bibfnamefont {B.}~\bibnamefont {Xi}}, \bibinfo {author} {\bibfnamefont
  {X.}~\bibnamefont {Shi}}, \bibinfo {author} {\bibfnamefont {F.}~\bibnamefont
  {Ye}}, \ and\ \bibinfo {author} {\bibfnamefont {J.-W.}\ \bibnamefont {Mei}},\
  }\href {\doibase 10.1103/PhysRevResearch.2.013214} {\bibfield  {journal}
  {\bibinfo  {journal} {Phys. Rev. Research}\ }\textbf {\bibinfo {volume}
  {2}},\ \bibinfo {pages} {013214} (\bibinfo {year}
  {2020}{\natexlab{a}})}\BibitemShut {NoStop}%
\bibitem [{\citenamefont {Jiang}\ \emph {et~al.}(2019)\citenamefont {Jiang},
  \citenamefont {Si}, \citenamefont {Liao},\ and\ \citenamefont
  {Zhong}}]{jiang2019electronic}%
  \BibitemOpen
  \bibfield  {author} {\bibinfo {author} {\bibfnamefont {P.}~\bibnamefont
  {Jiang}}, \bibinfo {author} {\bibfnamefont {L.}~\bibnamefont {Si}}, \bibinfo
  {author} {\bibfnamefont {Z.}~\bibnamefont {Liao}}, \ and\ \bibinfo {author}
  {\bibfnamefont {Z.}~\bibnamefont {Zhong}},\ }\href {\doibase
  10.1103/PhysRevB.100.201106} {\bibfield  {journal} {\bibinfo  {journal}
  {Phys. Rev. B}\ }\textbf {\bibinfo {volume} {100}},\ \bibinfo {pages}
  {201106} (\bibinfo {year} {2019})}\BibitemShut {NoStop}%
\bibitem [{\citenamefont {Hirayama}\ \emph {et~al.}(2020)\citenamefont
  {Hirayama}, \citenamefont {Tadano}, \citenamefont {Nomura},\ and\
  \citenamefont {Arita}}]{hirayama2019materials}%
  \BibitemOpen
  \bibfield  {author} {\bibinfo {author} {\bibfnamefont {M.}~\bibnamefont
  {Hirayama}}, \bibinfo {author} {\bibfnamefont {T.}~\bibnamefont {Tadano}},
  \bibinfo {author} {\bibfnamefont {Y.}~\bibnamefont {Nomura}}, \ and\ \bibinfo
  {author} {\bibfnamefont {R.}~\bibnamefont {Arita}},\ }\href {\doibase
  10.1103/PhysRevB.101.075107} {\bibfield  {journal} {\bibinfo  {journal}
  {Phys. Rev. B}\ }\textbf {\bibinfo {volume} {101}},\ \bibinfo {pages}
  {075107} (\bibinfo {year} {2020})}\BibitemShut {NoStop}%
\bibitem [{\citenamefont {Lechermann}(2020)}]{lechermann2019late}%
  \BibitemOpen
  \bibfield  {author} {\bibinfo {author} {\bibfnamefont {F.}~\bibnamefont
  {Lechermann}},\ }\href {\doibase 10.1103/PhysRevB.101.081110} {\bibfield
  {journal} {\bibinfo  {journal} {Phys. Rev. B}\ }\textbf {\bibinfo {volume}
  {101}},\ \bibinfo {pages} {081110} (\bibinfo {year} {2020})}\BibitemShut
  {NoStop}%
\bibitem [{\citenamefont {Gu}\ \emph {et~al.}()\citenamefont {Gu},
  \citenamefont {Zhu}, \citenamefont {Wang}, \citenamefont {Hu},\ and\
  \citenamefont {Chen}}]{gu2019}%
  \BibitemOpen
  \bibfield  {author} {\bibinfo {author} {\bibfnamefont {Y.}~\bibnamefont
  {Gu}}, \bibinfo {author} {\bibfnamefont {S.}~\bibnamefont {Zhu}}, \bibinfo
  {author} {\bibfnamefont {X.}~\bibnamefont {Wang}}, \bibinfo {author}
  {\bibfnamefont {J.}~\bibnamefont {Hu}}, \ and\ \bibinfo {author}
  {\bibfnamefont {H.}~\bibnamefont {Chen}},\ }\href@noop {} {\ }\Eprint
  {http://arxiv.org/abs/1911.00814} {arXiv:1911.00814} \BibitemShut {NoStop}%
\bibitem [{\citenamefont {Ryee}\ \emph {et~al.}(2020)\citenamefont {Ryee},
  \citenamefont {Yoon}, \citenamefont {Kim}, \citenamefont {Jeong},\ and\
  \citenamefont {Han}}]{ryee2019induced}%
  \BibitemOpen
  \bibfield  {author} {\bibinfo {author} {\bibfnamefont {S.}~\bibnamefont
  {Ryee}}, \bibinfo {author} {\bibfnamefont {H.}~\bibnamefont {Yoon}}, \bibinfo
  {author} {\bibfnamefont {T.~J.}\ \bibnamefont {Kim}}, \bibinfo {author}
  {\bibfnamefont {M.~Y.}\ \bibnamefont {Jeong}}, \ and\ \bibinfo {author}
  {\bibfnamefont {M.~J.}\ \bibnamefont {Han}},\ }\href {\doibase
  10.1103/PhysRevB.101.064513} {\bibfield  {journal} {\bibinfo  {journal}
  {Phys. Rev. B}\ }\textbf {\bibinfo {volume} {101}},\ \bibinfo {pages}
  {064513} (\bibinfo {year} {2020})}\BibitemShut {NoStop}%
\bibitem [{\citenamefont {Si}\ \emph {et~al.}()\citenamefont {Si},
  \citenamefont {Xiao}, \citenamefont {Kaufmann}, \citenamefont {Tomczak},
  \citenamefont {Lu}, \citenamefont {Zhong},\ and\ \citenamefont
  {Held}}]{si2019topotactic}%
  \BibitemOpen
  \bibfield  {author} {\bibinfo {author} {\bibfnamefont {L.}~\bibnamefont
  {Si}}, \bibinfo {author} {\bibfnamefont {W.}~\bibnamefont {Xiao}}, \bibinfo
  {author} {\bibfnamefont {J.}~\bibnamefont {Kaufmann}}, \bibinfo {author}
  {\bibfnamefont {J.~M.}\ \bibnamefont {Tomczak}}, \bibinfo {author}
  {\bibfnamefont {Y.}~\bibnamefont {Lu}}, \bibinfo {author} {\bibfnamefont
  {Z.}~\bibnamefont {Zhong}}, \ and\ \bibinfo {author} {\bibfnamefont
  {K.}~\bibnamefont {Held}},\ }\href@noop {} {\ }\Eprint
  {http://arxiv.org/abs/1911.06917} {arXiv:1911.06917} \BibitemShut {NoStop}%
\bibitem [{\citenamefont {Choi}\ \emph {et~al.}(2020)\citenamefont {Choi},
  \citenamefont {Lee},\ and\ \citenamefont {Pickett}}]{choi2019role}%
  \BibitemOpen
  \bibfield  {author} {\bibinfo {author} {\bibfnamefont {M.-Y.}\ \bibnamefont
  {Choi}}, \bibinfo {author} {\bibfnamefont {K.-W.}\ \bibnamefont {Lee}}, \
  and\ \bibinfo {author} {\bibfnamefont {W.~E.}\ \bibnamefont {Pickett}},\
  }\href {\doibase 10.1103/PhysRevB.101.020503} {\bibfield  {journal} {\bibinfo
   {journal} {Phys. Rev. B}\ }\textbf {\bibinfo {volume} {101}},\ \bibinfo
  {pages} {020503} (\bibinfo {year} {2020})}\BibitemShut {NoStop}%
\bibitem [{\citenamefont {Liu}\ \emph {et~al.}()\citenamefont {Liu},
  \citenamefont {Ren}, \citenamefont {Zhu}, \citenamefont {Wang},\ and\
  \citenamefont {Yang}}]{liu2019electronic}%
  \BibitemOpen
  \bibfield  {author} {\bibinfo {author} {\bibfnamefont {Z.}~\bibnamefont
  {Liu}}, \bibinfo {author} {\bibfnamefont {Z.}~\bibnamefont {Ren}}, \bibinfo
  {author} {\bibfnamefont {W.}~\bibnamefont {Zhu}}, \bibinfo {author}
  {\bibfnamefont {Z.~F.}\ \bibnamefont {Wang}}, \ and\ \bibinfo {author}
  {\bibfnamefont {J.}~\bibnamefont {Yang}},\ }\href@noop {} {\ }\Eprint
  {http://arxiv.org/abs/1912.01332} {arXiv:1912.01332} \BibitemShut {NoStop}%
\bibitem [{\citenamefont {Werner}\ and\ \citenamefont
  {Hoshino}(2020)}]{werner2019nickelate}%
  \BibitemOpen
  \bibfield  {author} {\bibinfo {author} {\bibfnamefont {P.}~\bibnamefont
  {Werner}}\ and\ \bibinfo {author} {\bibfnamefont {S.}~\bibnamefont
  {Hoshino}},\ }\href {\doibase 10.1103/PhysRevB.101.041104} {\bibfield
  {journal} {\bibinfo  {journal} {Phys. Rev. B}\ }\textbf {\bibinfo {volume}
  {101}},\ \bibinfo {pages} {041104} (\bibinfo {year} {2020})}\BibitemShut
  {NoStop}%
\bibitem [{\citenamefont {Jiang}\ \emph {et~al.}()\citenamefont {Jiang},
  \citenamefont {Berciu},\ and\ \citenamefont {Sawatzky}}]{jiang2019}%
  \BibitemOpen
  \bibfield  {author} {\bibinfo {author} {\bibfnamefont {M.}~\bibnamefont
  {Jiang}}, \bibinfo {author} {\bibfnamefont {M.}~\bibnamefont {Berciu}}, \
  and\ \bibinfo {author} {\bibfnamefont {G.~A.}\ \bibnamefont {Sawatzky}},\
  }\href@noop {} {\ }\Eprint {http://arxiv.org/abs/1909.02557}
  {arXiv:1909.02557} \BibitemShut {NoStop}%
\bibitem [{\citenamefont {Hirsch}\ and\ \citenamefont
  {Marsiglio}(2019)}]{Hirsch2019}%
  \BibitemOpen
  \bibfield  {author} {\bibinfo {author} {\bibfnamefont {J.}~\bibnamefont
  {Hirsch}}\ and\ \bibinfo {author} {\bibfnamefont {F.}~\bibnamefont
  {Marsiglio}},\ }\href {\doibase https://doi.org/10.1016/j.physc.2019.1353534}
  {\bibfield  {journal} {\bibinfo  {journal} {Physica C}\ }\textbf {\bibinfo
  {volume} {566}},\ \bibinfo {pages} {1353534} (\bibinfo {year}
  {2019})}\BibitemShut {NoStop}%
\bibitem [{\citenamefont {Zhang}\ \emph
  {et~al.}(2020{\natexlab{b}})\citenamefont {Zhang}, \citenamefont {Yang},\
  and\ \citenamefont {Zhang}}]{zhang2019selfdoped}%
  \BibitemOpen
  \bibfield  {author} {\bibinfo {author} {\bibfnamefont {G.}~\bibnamefont
  {Zhang}}, \bibinfo {author} {\bibfnamefont {Y.}~\bibnamefont {Yang}}, \ and\
  \bibinfo {author} {\bibfnamefont {F.}~\bibnamefont {Zhang}},\ }\href
  {\doibase 10.1103/PhysRevB.101.020501} {\bibfield  {journal} {\bibinfo
  {journal} {Phys. Rev. B}\ }\textbf {\bibinfo {volume} {101}},\ \bibinfo
  {pages} {020501} (\bibinfo {year} {2020}{\natexlab{b}})}\BibitemShut
  {NoStop}%
\bibitem [{\citenamefont {Zhang}\ and\ \citenamefont
  {Vishwanath}()}]{zhang2019kondo}%
  \BibitemOpen
  \bibfield  {author} {\bibinfo {author} {\bibfnamefont {Y.-H.}\ \bibnamefont
  {Zhang}}\ and\ \bibinfo {author} {\bibfnamefont {A.}~\bibnamefont
  {Vishwanath}},\ }\href@noop {} {\ }\Eprint {http://arxiv.org/abs/1909.12865}
  {arXiv:1909.12865} \BibitemShut {NoStop}%
\bibitem [{\citenamefont {Hu}\ and\ \citenamefont {Wu}(2019)}]{hu2019twoband}%
  \BibitemOpen
  \bibfield  {author} {\bibinfo {author} {\bibfnamefont {L.-H.}\ \bibnamefont
  {Hu}}\ and\ \bibinfo {author} {\bibfnamefont {C.}~\bibnamefont {Wu}},\ }\href
  {\doibase 10.1103/PhysRevResearch.1.032046} {\bibfield  {journal} {\bibinfo
  {journal} {Phys. Rev. Research}\ }\textbf {\bibinfo {volume} {1}},\ \bibinfo
  {pages} {032046} (\bibinfo {year} {2019})}\BibitemShut {NoStop}%
\bibitem [{\citenamefont {Eskes}\ and\ \citenamefont
  {Sawatzky}(1988)}]{Eskes88}%
  \BibitemOpen
  \bibfield  {author} {\bibinfo {author} {\bibfnamefont {H.}~\bibnamefont
  {Eskes}}\ and\ \bibinfo {author} {\bibfnamefont {G.~A.}\ \bibnamefont
  {Sawatzky}},\ }\href {\doibase 10.1103/PhysRevLett.61.1415} {\bibfield
  {journal} {\bibinfo  {journal} {Phys. Rev. Lett.}\ }\textbf {\bibinfo
  {volume} {61}},\ \bibinfo {pages} {1415} (\bibinfo {year}
  {1988})}\BibitemShut {NoStop}%
\bibitem [{\citenamefont {Georges}\ \emph {et~al.}(1996)\citenamefont
  {Georges}, \citenamefont {Kotliar}, \citenamefont {Krauth},\ and\
  \citenamefont {Rozenberg}}]{Georges1996}%
  \BibitemOpen
  \bibfield  {author} {\bibinfo {author} {\bibfnamefont {A.}~\bibnamefont
  {Georges}}, \bibinfo {author} {\bibfnamefont {G.}~\bibnamefont {Kotliar}},
  \bibinfo {author} {\bibfnamefont {W.}~\bibnamefont {Krauth}}, \ and\ \bibinfo
  {author} {\bibfnamefont {M.~J.}\ \bibnamefont {Rozenberg}},\ }\href {\doibase
  10.1103/RevModPhys.68.13} {\bibfield  {journal} {\bibinfo  {journal} {Rev.
  Mod. Phys.}\ }\textbf {\bibinfo {volume} {68}},\ \bibinfo {pages} {13}
  (\bibinfo {year} {1996})}\BibitemShut {NoStop}%
\bibitem [{\citenamefont {Georges}(2004)}]{Georges04}%
  \BibitemOpen
  \bibfield  {author} {\bibinfo {author} {\bibfnamefont {A.}~\bibnamefont
  {Georges}},\ }\href {\doibase 10.1063/1.1800733} {\bibfield  {journal}
  {\bibinfo  {journal} {AIP Conference Proceedings}\ }\textbf {\bibinfo
  {volume} {715}},\ \bibinfo {pages} {3} (\bibinfo {year} {2004})}\BibitemShut
  {NoStop}%
\bibitem [{\citenamefont {Kotliar}\ \emph {et~al.}(2006)\citenamefont
  {Kotliar}, \citenamefont {Savrasov}, \citenamefont {Haule}, \citenamefont
  {Oudovenko}, \citenamefont {Parcollet},\ and\ \citenamefont
  {Marianetti}}]{Kotliar06}%
  \BibitemOpen
  \bibfield  {author} {\bibinfo {author} {\bibfnamefont {G.}~\bibnamefont
  {Kotliar}}, \bibinfo {author} {\bibfnamefont {S.~Y.}\ \bibnamefont
  {Savrasov}}, \bibinfo {author} {\bibfnamefont {K.}~\bibnamefont {Haule}},
  \bibinfo {author} {\bibfnamefont {V.~S.}\ \bibnamefont {Oudovenko}}, \bibinfo
  {author} {\bibfnamefont {O.}~\bibnamefont {Parcollet}}, \ and\ \bibinfo
  {author} {\bibfnamefont {C.~A.}\ \bibnamefont {Marianetti}},\ }\href
  {\doibase 10.1103/RevModPhys.78.865} {\bibfield  {journal} {\bibinfo
  {journal} {Rev. Mod. Phys.}\ }\textbf {\bibinfo {volume} {78}},\ \bibinfo
  {pages} {865} (\bibinfo {year} {2006})}\BibitemShut {NoStop}%
\bibitem [{\citenamefont {Held}\ \emph {et~al.}(2006)\citenamefont {Held},
  \citenamefont {Nekrasov}, \citenamefont {Keller}, \citenamefont {Eyert},
  \citenamefont {Blümer}, \citenamefont {McMahan}, \citenamefont {Scalettar},
  \citenamefont {Pruschke}, \citenamefont {Anisimov},\ and\ \citenamefont
  {Vollhardt}}]{Held06}%
  \BibitemOpen
  \bibfield  {author} {\bibinfo {author} {\bibfnamefont {K.}~\bibnamefont
  {Held}}, \bibinfo {author} {\bibfnamefont {I.~A.}\ \bibnamefont {Nekrasov}},
  \bibinfo {author} {\bibfnamefont {G.}~\bibnamefont {Keller}}, \bibinfo
  {author} {\bibfnamefont {V.}~\bibnamefont {Eyert}}, \bibinfo {author}
  {\bibfnamefont {N.}~\bibnamefont {Blümer}}, \bibinfo {author} {\bibfnamefont
  {A.~K.}\ \bibnamefont {McMahan}}, \bibinfo {author} {\bibfnamefont {R.~T.}\
  \bibnamefont {Scalettar}}, \bibinfo {author} {\bibfnamefont {T.}~\bibnamefont
  {Pruschke}}, \bibinfo {author} {\bibfnamefont {V.~I.}\ \bibnamefont
  {Anisimov}}, \ and\ \bibinfo {author} {\bibfnamefont {D.}~\bibnamefont
  {Vollhardt}},\ }\href {\doibase 10.1002/pssb.200642053} {\bibfield  {journal}
  {\bibinfo  {journal} {Phys. Stat. Sol. (b)}\ }\textbf {\bibinfo {volume}
  {243}},\ \bibinfo {pages} {2599} (\bibinfo {year} {2006})}\BibitemShut
  {NoStop}%
\bibitem [{\citenamefont {Giannozzi}\ \emph {et~al.}(2009)\citenamefont
  {Giannozzi}, \citenamefont {Baroni}, \citenamefont {Bonini}, \citenamefont
  {Calandra}, \citenamefont {Car}, \citenamefont {Cavazzoni}, \citenamefont
  {Ceresoli}, \citenamefont {Chiarotti}, \citenamefont {Cococcioni},
  \citenamefont {Dabo}, \citenamefont {Corso}, \citenamefont {Fabris},
  \citenamefont {Fratesi}, \citenamefont {de~Gironcoli}, \citenamefont
  {Gebauer}, \citenamefont {Gerstmann}, \citenamefont {Gougoussis},
  \citenamefont {Kokalj}, \citenamefont {Lazzeri}, \citenamefont
  {Martin-Samos}, \citenamefont {Marzari}, \citenamefont {Mauri}, \citenamefont
  {Mazzarello}, \citenamefont {Paolini}, \citenamefont {Pasquarello},
  \citenamefont {Paulatto}, \citenamefont {Sbraccia}, \citenamefont {Scandolo},
  \citenamefont {Sclauzero}, \citenamefont {Seitsonen}, \citenamefont
  {Smogunov}, \citenamefont {Umari},\ and\ \citenamefont {Wentzcovitch}}]{QE}%
  \BibitemOpen
  \bibfield  {author} {\bibinfo {author} {\bibfnamefont {P.}~\bibnamefont
  {Giannozzi}}, \bibinfo {author} {\bibfnamefont {S.}~\bibnamefont {Baroni}},
  \bibinfo {author} {\bibfnamefont {N.}~\bibnamefont {Bonini}}, \bibinfo
  {author} {\bibfnamefont {M.}~\bibnamefont {Calandra}}, \bibinfo {author}
  {\bibfnamefont {R.}~\bibnamefont {Car}}, \bibinfo {author} {\bibfnamefont
  {C.}~\bibnamefont {Cavazzoni}}, \bibinfo {author} {\bibfnamefont
  {D.}~\bibnamefont {Ceresoli}}, \bibinfo {author} {\bibfnamefont {G.~L.}\
  \bibnamefont {Chiarotti}}, \bibinfo {author} {\bibfnamefont {M.}~\bibnamefont
  {Cococcioni}}, \bibinfo {author} {\bibfnamefont {I.}~\bibnamefont {Dabo}},
  \bibinfo {author} {\bibfnamefont {A.~D.}\ \bibnamefont {Corso}}, \bibinfo
  {author} {\bibfnamefont {S.}~\bibnamefont {Fabris}}, \bibinfo {author}
  {\bibfnamefont {G.}~\bibnamefont {Fratesi}}, \bibinfo {author} {\bibfnamefont
  {S.}~\bibnamefont {de~Gironcoli}}, \bibinfo {author} {\bibfnamefont
  {R.}~\bibnamefont {Gebauer}}, \bibinfo {author} {\bibfnamefont
  {U.}~\bibnamefont {Gerstmann}}, \bibinfo {author} {\bibfnamefont
  {C.}~\bibnamefont {Gougoussis}}, \bibinfo {author} {\bibfnamefont
  {A.}~\bibnamefont {Kokalj}}, \bibinfo {author} {\bibfnamefont
  {M.}~\bibnamefont {Lazzeri}}, \bibinfo {author} {\bibfnamefont
  {L.}~\bibnamefont {Martin-Samos}}, \bibinfo {author} {\bibfnamefont
  {N.}~\bibnamefont {Marzari}}, \bibinfo {author} {\bibfnamefont
  {F.}~\bibnamefont {Mauri}}, \bibinfo {author} {\bibfnamefont
  {R.}~\bibnamefont {Mazzarello}}, \bibinfo {author} {\bibfnamefont
  {S.}~\bibnamefont {Paolini}}, \bibinfo {author} {\bibfnamefont
  {A.}~\bibnamefont {Pasquarello}}, \bibinfo {author} {\bibfnamefont
  {L.}~\bibnamefont {Paulatto}}, \bibinfo {author} {\bibfnamefont
  {C.}~\bibnamefont {Sbraccia}}, \bibinfo {author} {\bibfnamefont
  {S.}~\bibnamefont {Scandolo}}, \bibinfo {author} {\bibfnamefont
  {G.}~\bibnamefont {Sclauzero}}, \bibinfo {author} {\bibfnamefont {A.~P.}\
  \bibnamefont {Seitsonen}}, \bibinfo {author} {\bibfnamefont {A.}~\bibnamefont
  {Smogunov}}, \bibinfo {author} {\bibfnamefont {P.}~\bibnamefont {Umari}}, \
  and\ \bibinfo {author} {\bibfnamefont {R.~M.}\ \bibnamefont {Wentzcovitch}},\
  }\href {\doibase 10.1088/0953-8984/21/39/395502} {\bibfield  {journal}
  {\bibinfo  {journal} {J. Phys. Condens. Matter}\ }\textbf {\bibinfo {volume}
  {21}},\ \bibinfo {pages} {395502} (\bibinfo {year} {2009})}\BibitemShut
  {NoStop}%
\bibitem [{\citenamefont {Perdew}\ \emph {et~al.}(1996)\citenamefont {Perdew},
  \citenamefont {Burke},\ and\ \citenamefont {Ernzerhof}}]{PBE}%
  \BibitemOpen
  \bibfield  {author} {\bibinfo {author} {\bibfnamefont {J.~P.}\ \bibnamefont
  {Perdew}}, \bibinfo {author} {\bibfnamefont {K.}~\bibnamefont {Burke}}, \
  and\ \bibinfo {author} {\bibfnamefont {M.}~\bibnamefont {Ernzerhof}},\ }\href
  {\doibase 10.1103/PhysRevLett.77.3865} {\bibfield  {journal} {\bibinfo
  {journal} {Phys. Rev. Lett.}\ }\textbf {\bibinfo {volume} {77}},\ \bibinfo
  {pages} {3865} (\bibinfo {year} {1996})}\BibitemShut {NoStop}%
\bibitem [{\citenamefont {Pizzi}\ \emph {et~al.}(2019)\citenamefont {Pizzi},
  \citenamefont {Vitale}, \citenamefont {Arita}, \citenamefont {Bluegel},
  \citenamefont {Freimuth}, \citenamefont {Géranton}, \citenamefont
  {Gibertini}, \citenamefont {Gresch}, \citenamefont {Johnson}, \citenamefont
  {Koretsune}, \citenamefont {Ibanez}, \citenamefont {Lee}, \citenamefont
  {Lihm}, \citenamefont {Marchand}, \citenamefont {Marrazzo}, \citenamefont
  {Mokrousov}, \citenamefont {Mustafa}, \citenamefont {Nohara}, \citenamefont
  {Nomura}, \citenamefont {Paulatto}, \citenamefont {Ponce}, \citenamefont
  {Ponweiser}, \citenamefont {Qiao}, \citenamefont {Thöle}, \citenamefont
  {Tsirkin}, \citenamefont {Wierzbowska}, \citenamefont {Marzari},
  \citenamefont {Vanderbilt}, \citenamefont {Souza}, \citenamefont {Mostofi},\
  and\ \citenamefont {Yates}}]{wannier90_v3}%
  \BibitemOpen
  \bibfield  {author} {\bibinfo {author} {\bibfnamefont {G.}~\bibnamefont
  {Pizzi}}, \bibinfo {author} {\bibfnamefont {V.}~\bibnamefont {Vitale}},
  \bibinfo {author} {\bibfnamefont {R.}~\bibnamefont {Arita}}, \bibinfo
  {author} {\bibfnamefont {S.}~\bibnamefont {Bluegel}}, \bibinfo {author}
  {\bibfnamefont {F.}~\bibnamefont {Freimuth}}, \bibinfo {author}
  {\bibfnamefont {G.}~\bibnamefont {Géranton}}, \bibinfo {author}
  {\bibfnamefont {M.}~\bibnamefont {Gibertini}}, \bibinfo {author}
  {\bibfnamefont {D.}~\bibnamefont {Gresch}}, \bibinfo {author} {\bibfnamefont
  {C.}~\bibnamefont {Johnson}}, \bibinfo {author} {\bibfnamefont
  {T.}~\bibnamefont {Koretsune}}, \bibinfo {author} {\bibfnamefont
  {J.}~\bibnamefont {Ibanez}}, \bibinfo {author} {\bibfnamefont
  {H.}~\bibnamefont {Lee}}, \bibinfo {author} {\bibfnamefont {J.-M.}\
  \bibnamefont {Lihm}}, \bibinfo {author} {\bibfnamefont {D.}~\bibnamefont
  {Marchand}}, \bibinfo {author} {\bibfnamefont {A.}~\bibnamefont {Marrazzo}},
  \bibinfo {author} {\bibfnamefont {Y.}~\bibnamefont {Mokrousov}}, \bibinfo
  {author} {\bibfnamefont {J.~I.}\ \bibnamefont {Mustafa}}, \bibinfo {author}
  {\bibfnamefont {Y.}~\bibnamefont {Nohara}}, \bibinfo {author} {\bibfnamefont
  {Y.}~\bibnamefont {Nomura}}, \bibinfo {author} {\bibfnamefont
  {L.}~\bibnamefont {Paulatto}}, \bibinfo {author} {\bibfnamefont
  {S.}~\bibnamefont {Ponce}}, \bibinfo {author} {\bibfnamefont
  {T.}~\bibnamefont {Ponweiser}}, \bibinfo {author} {\bibfnamefont
  {J.}~\bibnamefont {Qiao}}, \bibinfo {author} {\bibfnamefont {F.}~\bibnamefont
  {Thöle}}, \bibinfo {author} {\bibfnamefont {S.~S.}\ \bibnamefont {Tsirkin}},
  \bibinfo {author} {\bibfnamefont {M.}~\bibnamefont {Wierzbowska}}, \bibinfo
  {author} {\bibfnamefont {N.}~\bibnamefont {Marzari}}, \bibinfo {author}
  {\bibfnamefont {D.}~\bibnamefont {Vanderbilt}}, \bibinfo {author}
  {\bibfnamefont {I.}~\bibnamefont {Souza}}, \bibinfo {author} {\bibfnamefont
  {A.~A.}\ \bibnamefont {Mostofi}}, \ and\ \bibinfo {author} {\bibfnamefont
  {J.~R.}\ \bibnamefont {Yates}},\ }\href
  {http://iopscience.iop.org/10.1088/1361-648X/ab51ff} {\bibfield  {journal}
  {\bibinfo  {journal} {J. Phys. Condens. Matter}\ } (\bibinfo {year}
  {2019})}\BibitemShut {NoStop}%
\bibitem [{\citenamefont {Marzari}\ and\ \citenamefont
  {Vanderbilt}(1997)}]{MLWF1}%
  \BibitemOpen
  \bibfield  {author} {\bibinfo {author} {\bibfnamefont {N.}~\bibnamefont
  {Marzari}}\ and\ \bibinfo {author} {\bibfnamefont {D.}~\bibnamefont
  {Vanderbilt}},\ }\href {\doibase 10.1103/PhysRevB.56.12847} {\bibfield
  {journal} {\bibinfo  {journal} {Phys. Rev. B}\ }\textbf {\bibinfo {volume}
  {56}},\ \bibinfo {pages} {12847} (\bibinfo {year} {1997})}\BibitemShut
  {NoStop}%
\bibitem [{\citenamefont {Souza}\ \emph {et~al.}(2001)\citenamefont {Souza},
  \citenamefont {Marzari},\ and\ \citenamefont {Vanderbilt}}]{MLWF2}%
  \BibitemOpen
  \bibfield  {author} {\bibinfo {author} {\bibfnamefont {I.}~\bibnamefont
  {Souza}}, \bibinfo {author} {\bibfnamefont {N.}~\bibnamefont {Marzari}}, \
  and\ \bibinfo {author} {\bibfnamefont {D.}~\bibnamefont {Vanderbilt}},\
  }\href {\doibase 10.1103/PhysRevB.65.035109} {\bibfield  {journal} {\bibinfo
  {journal} {Phys. Rev. B}\ }\textbf {\bibinfo {volume} {65}},\ \bibinfo
  {pages} {035109} (\bibinfo {year} {2001})}\BibitemShut {NoStop}%
\bibitem [{\citenamefont {Wang}\ \emph {et~al.}(2014)\citenamefont {Wang},
  \citenamefont {Lazar}, \citenamefont {Park}, \citenamefont {Millis},\ and\
  \citenamefont {Marianetti}}]{Wang14}%
  \BibitemOpen
  \bibfield  {author} {\bibinfo {author} {\bibfnamefont {R.}~\bibnamefont
  {Wang}}, \bibinfo {author} {\bibfnamefont {E.~A.}\ \bibnamefont {Lazar}},
  \bibinfo {author} {\bibfnamefont {H.}~\bibnamefont {Park}}, \bibinfo {author}
  {\bibfnamefont {A.~J.}\ \bibnamefont {Millis}}, \ and\ \bibinfo {author}
  {\bibfnamefont {C.~A.}\ \bibnamefont {Marianetti}},\ }\href {\doibase
  10.1103/PhysRevB.90.165125} {\bibfield  {journal} {\bibinfo  {journal} {Phys.
  Rev. B}\ }\textbf {\bibinfo {volume} {90}},\ \bibinfo {pages} {165125}
  (\bibinfo {year} {2014})}\BibitemShut {NoStop}%
\bibitem [{\citenamefont {Kanamori}(1963)}]{Kanamori1963}%
  \BibitemOpen
  \bibfield  {author} {\bibinfo {author} {\bibfnamefont {J.}~\bibnamefont
  {Kanamori}},\ }\href {\doibase 10.1143/ptp.30.275} {\bibfield  {journal}
  {\bibinfo  {journal} {Progr. Theor. Phys}\ }\textbf {\bibinfo {volume}
  {30}},\ \bibinfo {pages} {275} (\bibinfo {year} {1963})}\BibitemShut
  {NoStop}%
\bibitem [{\citenamefont {Nowadnick}\ \emph {et~al.}(2015)\citenamefont
  {Nowadnick}, \citenamefont {Ruf}, \citenamefont {Park}, \citenamefont {King},
  \citenamefont {Schlom}, \citenamefont {Shen},\ and\ \citenamefont
  {Millis}}]{Nowadnick15}%
  \BibitemOpen
  \bibfield  {author} {\bibinfo {author} {\bibfnamefont {E.~A.}\ \bibnamefont
  {Nowadnick}}, \bibinfo {author} {\bibfnamefont {J.~P.}\ \bibnamefont {Ruf}},
  \bibinfo {author} {\bibfnamefont {H.}~\bibnamefont {Park}}, \bibinfo {author}
  {\bibfnamefont {P.~D.~C.}\ \bibnamefont {King}}, \bibinfo {author}
  {\bibfnamefont {D.~G.}\ \bibnamefont {Schlom}}, \bibinfo {author}
  {\bibfnamefont {K.~M.}\ \bibnamefont {Shen}}, \ and\ \bibinfo {author}
  {\bibfnamefont {A.~J.}\ \bibnamefont {Millis}},\ }\href {\doibase
  10.1103/PhysRevB.92.245109} {\bibfield  {journal} {\bibinfo  {journal} {Phys.
  Rev. B}\ }\textbf {\bibinfo {volume} {92}},\ \bibinfo {pages} {245109}
  (\bibinfo {year} {2015})}\BibitemShut {NoStop}%
\bibitem [{\citenamefont {Held}(2007)}]{Held2007Electronic}%
  \BibitemOpen
  \bibfield  {author} {\bibinfo {author} {\bibfnamefont {K.}~\bibnamefont
  {Held}},\ }\href {\doibase 10.1080/00018730701619647} {\bibfield  {journal}
  {\bibinfo  {journal} {Adv. Phys.}\ }\textbf {\bibinfo {volume} {56}},\
  \bibinfo {pages} {829} (\bibinfo {year} {2007})}\BibitemShut {NoStop}%
\bibitem [{\citenamefont {Parcollet}\ \emph {et~al.}(2015)\citenamefont
  {Parcollet}, \citenamefont {Ferrero}, \citenamefont {Ayral}, \citenamefont
  {Hafermann}, \citenamefont {Krivenko}, \citenamefont {Messio},\ and\
  \citenamefont {Seth}}]{TRIQS}%
  \BibitemOpen
  \bibfield  {author} {\bibinfo {author} {\bibfnamefont {O.}~\bibnamefont
  {Parcollet}}, \bibinfo {author} {\bibfnamefont {M.}~\bibnamefont {Ferrero}},
  \bibinfo {author} {\bibfnamefont {T.}~\bibnamefont {Ayral}}, \bibinfo
  {author} {\bibfnamefont {H.}~\bibnamefont {Hafermann}}, \bibinfo {author}
  {\bibfnamefont {I.}~\bibnamefont {Krivenko}}, \bibinfo {author}
  {\bibfnamefont {L.}~\bibnamefont {Messio}}, \ and\ \bibinfo {author}
  {\bibfnamefont {P.}~\bibnamefont {Seth}},\ }\href {\doibase
  http://dx.doi.org/10.1016/j.cpc.2015.04.023} {\bibfield  {journal} {\bibinfo
  {journal} {Comput. Phys. Commun.}\ }\textbf {\bibinfo {volume} {196}},\
  \bibinfo {pages} {398 } (\bibinfo {year} {2015})}\BibitemShut {NoStop}%
\bibitem [{\citenamefont {Aichhorn}\ \emph {et~al.}(2016)\citenamefont
  {Aichhorn}, \citenamefont {Pourovskii}, \citenamefont {Seth}, \citenamefont
  {Vildosola}, \citenamefont {Zingl}, \citenamefont {Peil}, \citenamefont
  {Deng}, \citenamefont {Mravlje}, \citenamefont {Kraberger}, \citenamefont
  {Martins}, \citenamefont {Ferrero},\ and\ \citenamefont
  {Parcollet}}]{TRIQS/DFTTOOLS}%
  \BibitemOpen
  \bibfield  {author} {\bibinfo {author} {\bibfnamefont {M.}~\bibnamefont
  {Aichhorn}}, \bibinfo {author} {\bibfnamefont {L.}~\bibnamefont
  {Pourovskii}}, \bibinfo {author} {\bibfnamefont {P.}~\bibnamefont {Seth}},
  \bibinfo {author} {\bibfnamefont {V.}~\bibnamefont {Vildosola}}, \bibinfo
  {author} {\bibfnamefont {M.}~\bibnamefont {Zingl}}, \bibinfo {author}
  {\bibfnamefont {O.~E.}\ \bibnamefont {Peil}}, \bibinfo {author}
  {\bibfnamefont {X.}~\bibnamefont {Deng}}, \bibinfo {author} {\bibfnamefont
  {J.}~\bibnamefont {Mravlje}}, \bibinfo {author} {\bibfnamefont {G.~J.}\
  \bibnamefont {Kraberger}}, \bibinfo {author} {\bibfnamefont {C.}~\bibnamefont
  {Martins}}, \bibinfo {author} {\bibfnamefont {M.}~\bibnamefont {Ferrero}}, \
  and\ \bibinfo {author} {\bibfnamefont {O.}~\bibnamefont {Parcollet}},\ }\href
  {\doibase 10.1016/j.cpc.2016.03.014} {\bibfield  {journal} {\bibinfo
  {journal} {Comput. Phys. Commun.}\ }\textbf {\bibinfo {volume} {204}},\
  \bibinfo {pages} {200 } (\bibinfo {year} {2016})}\BibitemShut {NoStop}%
\bibitem [{\citenamefont {Seth}\ \emph {et~al.}(2016)\citenamefont {Seth},
  \citenamefont {Krivenko}, \citenamefont {Ferrero},\ and\ \citenamefont
  {Parcollet}}]{TRIQS/CTHYB}%
  \BibitemOpen
  \bibfield  {author} {\bibinfo {author} {\bibfnamefont {P.}~\bibnamefont
  {Seth}}, \bibinfo {author} {\bibfnamefont {I.}~\bibnamefont {Krivenko}},
  \bibinfo {author} {\bibfnamefont {M.}~\bibnamefont {Ferrero}}, \ and\
  \bibinfo {author} {\bibfnamefont {O.}~\bibnamefont {Parcollet}},\ }\href
  {http://www.sciencedirect.com/science/article/pii/S001046551500404X}
  {\bibfield  {journal} {\bibinfo  {journal} {Comput. Phys. Commun.}\ }\textbf
  {\bibinfo {volume} {200}},\ \bibinfo {pages} {274 } (\bibinfo {year}
  {2016})}\BibitemShut {NoStop}%
\bibitem [{\citenamefont {Haule}(2007)}]{Haule_ctqmc}%
  \BibitemOpen
  \bibfield  {author} {\bibinfo {author} {\bibfnamefont {K.}~\bibnamefont
  {Haule}},\ }\href {\doibase 10.1103/PhysRevB.75.155113} {\bibfield  {journal}
  {\bibinfo  {journal} {Phys. Rev. B}\ }\textbf {\bibinfo {volume} {75}},\
  \bibinfo {pages} {155113} (\bibinfo {year} {2007})}\BibitemShut {NoStop}%
\bibitem [{\citenamefont {Park}\ \emph {et~al.}(2014)\citenamefont {Park},
  \citenamefont {Millis},\ and\ \citenamefont
  {Marianetti}}]{park2014computing}%
  \BibitemOpen
  \bibfield  {author} {\bibinfo {author} {\bibfnamefont {H.}~\bibnamefont
  {Park}}, \bibinfo {author} {\bibfnamefont {A.~J.}\ \bibnamefont {Millis}}, \
  and\ \bibinfo {author} {\bibfnamefont {C.~A.}\ \bibnamefont {Marianetti}},\
  }\href {\doibase 10.1103/PhysRevB.90.235103} {\bibfield  {journal} {\bibinfo
  {journal} {Phys. Rev. B}\ }\textbf {\bibinfo {volume} {90}},\ \bibinfo
  {pages} {235103} (\bibinfo {year} {2014})}\BibitemShut {NoStop}%
\bibitem [{\citenamefont {Chen}\ \emph {et~al.}(2015)\citenamefont {Chen},
  \citenamefont {Millis},\ and\ \citenamefont {Marianetti}}]{Chen2015Density}%
  \BibitemOpen
  \bibfield  {author} {\bibinfo {author} {\bibfnamefont {J.}~\bibnamefont
  {Chen}}, \bibinfo {author} {\bibfnamefont {A.~J.}\ \bibnamefont {Millis}}, \
  and\ \bibinfo {author} {\bibfnamefont {C.~A.}\ \bibnamefont {Marianetti}},\
  }\href {\doibase 10.1103/PhysRevB.91.241111} {\bibfield  {journal} {\bibinfo
  {journal} {Phys. Rev. B}\ }\textbf {\bibinfo {volume} {91}},\ \bibinfo
  {pages} {241111} (\bibinfo {year} {2015})}\BibitemShut {NoStop}%
\bibitem [{\citenamefont {Chen}\ and\ \citenamefont {Millis}(2016)}]{Chen16}%
  \BibitemOpen
  \bibfield  {author} {\bibinfo {author} {\bibfnamefont {H.}~\bibnamefont
  {Chen}}\ and\ \bibinfo {author} {\bibfnamefont {A.~J.}\ \bibnamefont
  {Millis}},\ }\href {\doibase 10.1103/PhysRevB.93.045133} {\bibfield
  {journal} {\bibinfo  {journal} {Phys. Rev. B}\ }\textbf {\bibinfo {volume}
  {93}},\ \bibinfo {pages} {045133} (\bibinfo {year} {2016})}\BibitemShut
  {NoStop}%
\bibitem [{\citenamefont {Liechtenstein}\ \emph {et~al.}(1995)\citenamefont
  {Liechtenstein}, \citenamefont {Anisimov},\ and\ \citenamefont
  {Zaanen}}]{liechtenstein1995fll}%
  \BibitemOpen
  \bibfield  {author} {\bibinfo {author} {\bibfnamefont {A.~I.}\ \bibnamefont
  {Liechtenstein}}, \bibinfo {author} {\bibfnamefont {V.~I.}\ \bibnamefont
  {Anisimov}}, \ and\ \bibinfo {author} {\bibfnamefont {J.}~\bibnamefont
  {Zaanen}},\ }\href {\doibase 10.1103/PhysRevB.52.R5467} {\bibfield  {journal}
  {\bibinfo  {journal} {Phys. Rev. B}\ }\textbf {\bibinfo {volume} {52}},\
  \bibinfo {pages} {R5467} (\bibinfo {year} {1995})}\BibitemShut {NoStop}%
\bibitem [{\citenamefont {Czy\ifmmode~\dot{z}\else \.{z}\fi{}yk}\ and\
  \citenamefont {Sawatzky}(1994)}]{czyzyk1994amf}%
  \BibitemOpen
  \bibfield  {author} {\bibinfo {author} {\bibfnamefont {M.~T.}\ \bibnamefont
  {Czy\ifmmode~\dot{z}\else \.{z}\fi{}yk}}\ and\ \bibinfo {author}
  {\bibfnamefont {G.~A.}\ \bibnamefont {Sawatzky}},\ }\href {\doibase
  10.1103/PhysRevB.49.14211} {\bibfield  {journal} {\bibinfo  {journal} {Phys.
  Rev. B}\ }\textbf {\bibinfo {volume} {49}},\ \bibinfo {pages} {14211}
  (\bibinfo {year} {1994})}\BibitemShut {NoStop}%
\bibitem [{\citenamefont {Dang}\ \emph {et~al.}(2014)\citenamefont {Dang},
  \citenamefont {Millis},\ and\ \citenamefont {Marianetti}}]{Dang14}%
  \BibitemOpen
  \bibfield  {author} {\bibinfo {author} {\bibfnamefont {H.~T.}\ \bibnamefont
  {Dang}}, \bibinfo {author} {\bibfnamefont {A.~J.}\ \bibnamefont {Millis}}, \
  and\ \bibinfo {author} {\bibfnamefont {C.~A.}\ \bibnamefont {Marianetti}},\
  }\href {\doibase 10.1103/PhysRevB.89.161113} {\bibfield  {journal} {\bibinfo
  {journal} {Phys. Rev. B}\ }\textbf {\bibinfo {volume} {89}},\ \bibinfo
  {pages} {161113} (\bibinfo {year} {2014})}\BibitemShut {NoStop}%
\bibitem [{\citenamefont {Haule}(2015)}]{Haule2015exact}%
  \BibitemOpen
  \bibfield  {author} {\bibinfo {author} {\bibfnamefont {K.}~\bibnamefont
  {Haule}},\ }\href {\doibase 10.1103/PhysRevLett.115.196403} {\bibfield
  {journal} {\bibinfo  {journal} {Phys. Rev. Lett.}\ }\textbf {\bibinfo
  {volume} {115}},\ \bibinfo {pages} {196403} (\bibinfo {year}
  {2015})}\BibitemShut {NoStop}%
\bibitem [{\citenamefont {Seth}\ \emph {et~al.}(2017)\citenamefont {Seth},
  \citenamefont {Peil}, \citenamefont {Pourovskii}, \citenamefont {Betzinger},
  \citenamefont {Friedrich}, \citenamefont {Parcollet}, \citenamefont
  {Biermann}, \citenamefont {Aryasetiawan},\ and\ \citenamefont
  {Georges}}]{Seth17}%
  \BibitemOpen
  \bibfield  {author} {\bibinfo {author} {\bibfnamefont {P.}~\bibnamefont
  {Seth}}, \bibinfo {author} {\bibfnamefont {O.~E.}\ \bibnamefont {Peil}},
  \bibinfo {author} {\bibfnamefont {L.}~\bibnamefont {Pourovskii}}, \bibinfo
  {author} {\bibfnamefont {M.}~\bibnamefont {Betzinger}}, \bibinfo {author}
  {\bibfnamefont {C.}~\bibnamefont {Friedrich}}, \bibinfo {author}
  {\bibfnamefont {O.}~\bibnamefont {Parcollet}}, \bibinfo {author}
  {\bibfnamefont {S.}~\bibnamefont {Biermann}}, \bibinfo {author}
  {\bibfnamefont {F.}~\bibnamefont {Aryasetiawan}}, \ and\ \bibinfo {author}
  {\bibfnamefont {A.}~\bibnamefont {Georges}},\ }\href {\doibase
  10.1103/PhysRevB.96.205139} {\bibfield  {journal} {\bibinfo  {journal} {Phys.
  Rev. B}\ }\textbf {\bibinfo {volume} {96}},\ \bibinfo {pages} {205139}
  (\bibinfo {year} {2017})}\BibitemShut {NoStop}%
\bibitem [{\citenamefont {Lichtenstein}\ \emph {et~al.}(2001)\citenamefont
  {Lichtenstein}, \citenamefont {Katsnelson},\ and\ \citenamefont
  {Kotliar}}]{Neel01}%
  \BibitemOpen
  \bibfield  {author} {\bibinfo {author} {\bibfnamefont {A.~I.}\ \bibnamefont
  {Lichtenstein}}, \bibinfo {author} {\bibfnamefont {M.~I.}\ \bibnamefont
  {Katsnelson}}, \ and\ \bibinfo {author} {\bibfnamefont {G.}~\bibnamefont
  {Kotliar}},\ }\href {\doibase 10.1103/PhysRevLett.87.067205} {\bibfield
  {journal} {\bibinfo  {journal} {Phys. Rev. Lett.}\ }\textbf {\bibinfo
  {volume} {87}},\ \bibinfo {pages} {067205} (\bibinfo {year}
  {2001})}\BibitemShut {NoStop}%
\bibitem [{\citenamefont {Mravlje}\ \emph {et~al.}(2012)\citenamefont
  {Mravlje}, \citenamefont {Aichhorn},\ and\ \citenamefont {Georges}}]{Neel02}%
  \BibitemOpen
  \bibfield  {author} {\bibinfo {author} {\bibfnamefont {J.}~\bibnamefont
  {Mravlje}}, \bibinfo {author} {\bibfnamefont {M.}~\bibnamefont {Aichhorn}}, \
  and\ \bibinfo {author} {\bibfnamefont {A.}~\bibnamefont {Georges}},\ }\href
  {\doibase 10.1103/PhysRevLett.108.197202} {\bibfield  {journal} {\bibinfo
  {journal} {Phys. Rev. Lett.}\ }\textbf {\bibinfo {volume} {108}},\ \bibinfo
  {pages} {197202} (\bibinfo {year} {2012})}\BibitemShut {NoStop}%
\bibitem [{\citenamefont {Karolak}\ \emph {et~al.}(2015)\citenamefont
  {Karolak}, \citenamefont {Edelmann},\ and\ \citenamefont
  {Sangiovanni}}]{Neel03}%
  \BibitemOpen
  \bibfield  {author} {\bibinfo {author} {\bibfnamefont {M.}~\bibnamefont
  {Karolak}}, \bibinfo {author} {\bibfnamefont {M.}~\bibnamefont {Edelmann}}, \
  and\ \bibinfo {author} {\bibfnamefont {G.}~\bibnamefont {Sangiovanni}},\
  }\href {\doibase 10.1103/PhysRevB.91.075108} {\bibfield  {journal} {\bibinfo
  {journal} {Phys. Rev. B}\ }\textbf {\bibinfo {volume} {91}},\ \bibinfo
  {pages} {075108} (\bibinfo {year} {2015})}\BibitemShut {NoStop}%
\bibitem [{\citenamefont {Zingl}\ \emph {et~al.}(2016)\citenamefont {Zingl},
  \citenamefont {Assmann}, \citenamefont {Seth}, \citenamefont {Krivenko},\
  and\ \citenamefont {Aichhorn}}]{Neel04}%
  \BibitemOpen
  \bibfield  {author} {\bibinfo {author} {\bibfnamefont {M.}~\bibnamefont
  {Zingl}}, \bibinfo {author} {\bibfnamefont {E.}~\bibnamefont {Assmann}},
  \bibinfo {author} {\bibfnamefont {P.}~\bibnamefont {Seth}}, \bibinfo {author}
  {\bibfnamefont {I.}~\bibnamefont {Krivenko}}, \ and\ \bibinfo {author}
  {\bibfnamefont {M.}~\bibnamefont {Aichhorn}},\ }\href {\doibase
  10.1103/PhysRevB.94.045130} {\bibfield  {journal} {\bibinfo  {journal} {Phys.
  Rev. B}\ }\textbf {\bibinfo {volume} {94}},\ \bibinfo {pages} {045130}
  (\bibinfo {year} {2016})}\BibitemShut {NoStop}%
\bibitem [{\citenamefont {Weber}\ \emph {et~al.}(2012)\citenamefont {Weber},
  \citenamefont {Yee}, \citenamefont {Haule},\ and\ \citenamefont
  {Kotliar}}]{Weber2012}%
  \BibitemOpen
  \bibfield  {author} {\bibinfo {author} {\bibfnamefont {C.}~\bibnamefont
  {Weber}}, \bibinfo {author} {\bibfnamefont {C.}~\bibnamefont {Yee}}, \bibinfo
  {author} {\bibfnamefont {K.}~\bibnamefont {Haule}}, \ and\ \bibinfo {author}
  {\bibfnamefont {G.}~\bibnamefont {Kotliar}},\ }\href {\doibase
  10.1209/0295-5075/100/37001} {\bibfield  {journal} {\bibinfo  {journal} {{EPL
  (Europhysics Letters)}}\ }\textbf {\bibinfo {volume} {100}},\ \bibinfo
  {pages} {37001} (\bibinfo {year} {2012})}\BibitemShut {NoStop}%
\bibitem [{Note1()}]{Note1}%
  \BibitemOpen
  \bibinfo {note} {We use: \protect \mbox {Nd.pbe-spdn-kjpaw\protect
  \_psl.1.0.0.UPF} \\ \protect \mbox {Ni.pbe-spn-kjpaw\protect
  \_psl.1.0.0.UPF,}\\ \protect \mbox {O .pbe-n-kjpaw\protect \_psl.1.0.0.UPF}
  \\ \protect \mbox {Ca.pbe-spn-kjpaw\protect \_psl.1.0.0.UPF,}\ \protect \mbox
  {Cu.pbe-dn-kjpaw\protect \_psl.1.0.0.UPF}}\BibitemShut {NoStop}%
\bibitem [{\citenamefont {Blaha}\ \emph {et~al.}(2018)\citenamefont {Blaha},
  \citenamefont {Schwarz}, \citenamefont {Madsen}, \citenamefont {Kvasnicka},
  \citenamefont {Luitz}, \citenamefont {Laskowski}, \citenamefont {Tran},\ and\
  \citenamefont {Marks}}]{Blaha2018}%
  \BibitemOpen
  \bibfield  {author} {\bibinfo {author} {\bibfnamefont {P.}~\bibnamefont
  {Blaha}}, \bibinfo {author} {\bibfnamefont {K.}~\bibnamefont {Schwarz}},
  \bibinfo {author} {\bibfnamefont {G.~K.~H.}\ \bibnamefont {Madsen}}, \bibinfo
  {author} {\bibfnamefont {D.}~\bibnamefont {Kvasnicka}}, \bibinfo {author}
  {\bibfnamefont {J.}~\bibnamefont {Luitz}}, \bibinfo {author} {\bibfnamefont
  {R.}~\bibnamefont {Laskowski}}, \bibinfo {author} {\bibfnamefont
  {F.}~\bibnamefont {Tran}}, \ and\ \bibinfo {author} {\bibfnamefont {L.~D.}\
  \bibnamefont {Marks}},\ }\href@noop {} {\emph {\bibinfo {title} {WIEN2k, An
  Augmented Plane Wave + Local Orbitals Program for Calculating Crystal
  Properties}}}\ (\bibinfo  {publisher} {K. Schwarz, Tech. Univ. Wien,
  Austria},\ \bibinfo {year} {2018})\BibitemShut {NoStop}%
\bibitem [{\citenamefont {Kune{\v s}}\ \emph {et~al.}(2010)\citenamefont
  {Kune{\v s}}, \citenamefont {Arita}, \citenamefont {Wissgott}, \citenamefont
  {Toschi}, \citenamefont {Ikeda},\ and\ \citenamefont {Held}}]{wien2wannier}%
  \BibitemOpen
  \bibfield  {author} {\bibinfo {author} {\bibfnamefont {J.}~\bibnamefont
  {Kune{\v s}}}, \bibinfo {author} {\bibfnamefont {R.}~\bibnamefont {Arita}},
  \bibinfo {author} {\bibfnamefont {P.}~\bibnamefont {Wissgott}}, \bibinfo
  {author} {\bibfnamefont {A.}~\bibnamefont {Toschi}}, \bibinfo {author}
  {\bibfnamefont {H.}~\bibnamefont {Ikeda}}, \ and\ \bibinfo {author}
  {\bibfnamefont {K.}~\bibnamefont {Held}},\ }\href {\doibase
  http://dx.doi.org/10.1016/j.cpc.2010.08.005} {\bibfield  {journal} {\bibinfo
  {journal} {Comput. Phys. Commun.}\ }\textbf {\bibinfo {volume} {181}},\
  \bibinfo {pages} {1888 } (\bibinfo {year} {2010})}\BibitemShut {NoStop}%
\bibitem [{\citenamefont {{TRIQS/maxent website}}()}]{TRIQS/maxent}%
  \BibitemOpen
  \bibfield  {author} {\bibinfo {author} {\bibnamefont {{TRIQS/maxent
  website}}},\ }\href@noop {} {\enquote {\bibinfo {title}
  {\url{https://triqs.github.io/maxent}},}\ }\BibitemShut {NoStop}%
\end{thebibliography}%

\newpage\section{Supplementary Information}
\subsection{DFT Calculations}

We use Quantum Espresso~\cite{QE} to perform the DFT calculations. We use PAW pseudopotentials~\footnote{We use: \mbox{Nd.pbe-spdn-kjpaw\_psl.1.0.0.UPF} \\ \mbox{Ni.pbe-spn-kjpaw\_psl.1.0.0.UPF,}\\ \mbox{O   .pbe-n-kjpaw\_psl.1.0.0.UPF} \\ \mbox{Ca.pbe-spn-kjpaw\_psl.1.0.0.UPF,}\ \mbox{Cu.pbe-dn-kjpaw\_psl.1.0.0.UPF}} with the Nd-$f$ states as part of the core. For both compounds we take the ideal tetragonal structure. For NdNiO$_2$ we use the structure parameters $a = \SI{3.95}{\angstrom}$ and $c = \SI{3.37}{\angstrom}$ and for CaCuO$_2$ we use $a = \SI{3.86}{\angstrom}$ and $c = \SI{3.20}{\angstrom}$. We employ the PBE-GGA functional~\cite{PBE}, a $k$-point mesh of $16 \times 16 \times 16$, an energy cutoff of \SI{70}{Ry} for the wavefunctions, and an energy cutoff of \SI{280}{Ry} for the density and potential. We get an identical band structure using a $k$-point mesh of $32 \times 32 \times 32$, a wavefunction energy cutoff of \SI{150}{Ry} and a density and potential cutoff of \SI{600}{Ry}. For CaCuO$_2$, we compared the band structure to an all electron calculation (WIEN2k~\cite{Blaha2018}) and find excellent agreement.

\subsection{Energy Windows and Wannierization}
The first step of the many-body calculations is to define a basis of single particle states, some of which are treated as correlated and some as uncorrelated or ``background'' states. The interaction parameters used for the correlated states depend on the basis chosen.  In DMFT calculations the single particle basis is typically defined by choosing an energy window and then defining the single particle states in terms of the DFT bands that fall within this window. For transition metal oxides, two choices of energy window are common and both are employed here. 

The ``wide energy window" is an energy range reaching $\sim 6$ to $\SI{10}{eV}$ below the chemical potential,  chosen to incorporate the O-$p$-derived states. Experience with other compounds indicates that the basis states constructed from the wide energy window are substantially similar to the corresponding  free ion states. The ``narrow energy window'' is an energy range that captures only the near Fermi level ``frontier'' orbitals, particularly the antibonding bands composed of  Cu/Ni $d_{x^2-y^2}$-O $p$-derived states. These states are physically rather more extended than the rest of the states defined in the wide energy window construction. In NdNiO$_2$ it is essential that the narrow energy window calculations also incorporate the Nd-derived bands that cross the Fermi level.

\newpage
To construct the single particle basis we construct maximally localized Wannier functions~\cite{MLWF1, MLWF2} using Wannier90~\cite{wien2wannier, wannier90_v3} and a dense $21 \times 21 \times 21$ $k$-point mesh.  The construction of the Nd-derived Wannier states has been the subject of discussion in the literature~\cite{nomura2019, hepting2019}. For NdNiO$_2$ in both the wide and narrow energy windows we use the ``selective localiztion'' method of Ref.~\cite{Wang14} to localize only  the Ni-d orbitals, which are also constrained to be centered on the Ni sites.  These restrictions improve convergence and provide a Hamiltonian which is to good accuracy real.

\begin{table}[t]
\begin{tabular}{|c|c|c|c|c|c|}
\hline
          &~TM $d_{x^2-y^2}$~&~TM $d_{3z^2-r^2}$~& ~~~~~O $p_\sigma$~~~~~ & ~Nd $d_{3z^2-r^2}$~ & ~~~~Nd $d_{xy}$~~~~ \\ \hline
~NdNiO$_2$~& -1.576              & -1.720          & -4.877     & 1.449        & 1.595       \\ \hline
~CaCuO$_2$~& -1.934              & -2.490          & -3.988     &      -       &          -  \\ \hline
\end{tabular}
\caption{On-site energies of the wide energy window Wannier model. TM stands for nickel or copper. The chemical potential is set to $0$.}
\label{tab:wide_onsite}
\end{table}

\begin{table}[t]
\begin{tabular}{|c|c|c|c|c|c|}
\hline
          & ~~TM $d_{x^2-y^2}$~~ & ~TM $d_{3z^2-r^2}$~& ~~~Nd $d_{xy}$~~~ &~TM $d_{x^2-y^2}$~& ~TM $d_{3z^2-r^2}$~ \\ 
          & O $p_\sigma$ & O $p_\sigma$ & O $p_\sigma$ & TM $d_{x^2-y^2}$ & TM $d_{3z^2-r^2}$ \\ \hline
~NdNiO$_2$~& 1.370                              & 0.396                          & 0.682                      & 0.131                                     & -0.084                            \\ \hline
~CaCuO$_2$~& 1.275                              & 0.257                          &                -           & -0.007                                    & -0.006                            \\ \hline
\end{tabular}
\caption{Hopping parameters of the wide energy window Wannier model. TM stands for nickel or copper.}
\label{tab:wide_hopping}
\end{table}

\begin{figure}[t]
    \centering
    \includegraphics[width = .48 \linewidth]{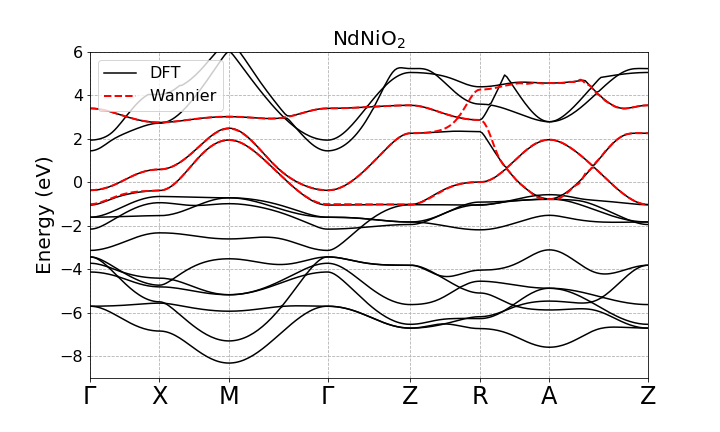}
    \includegraphics[width = .48 \linewidth]{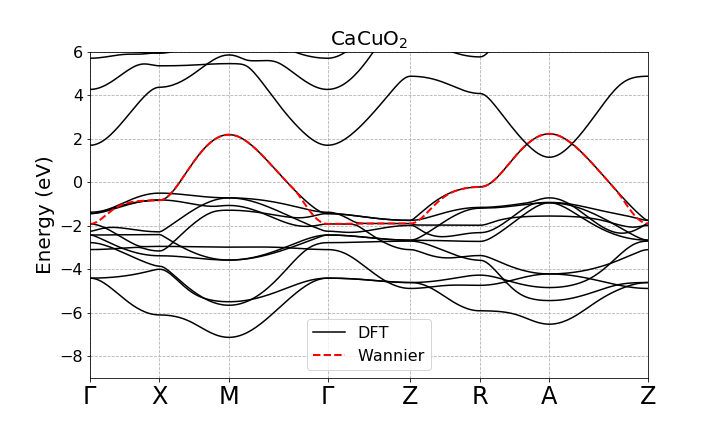}
    \caption{Wannier fits in the narrow energy window.}
    \label{fig:narrow_wannier}
\end{figure}

\begin{figure}[t]
    \centering
    \includegraphics[width = .48 \linewidth]{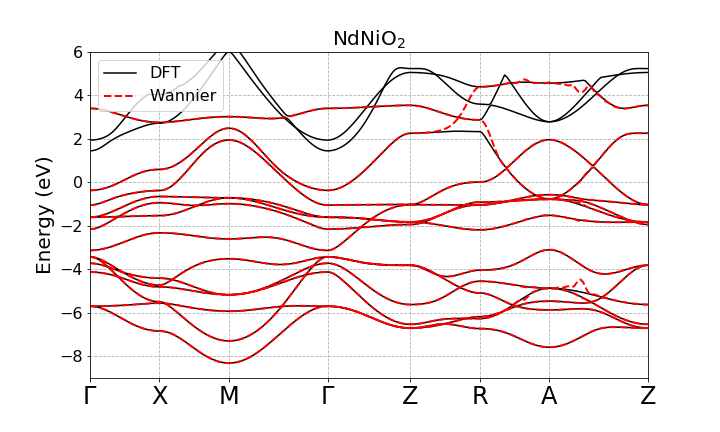}
    \includegraphics[width = .48 \linewidth]{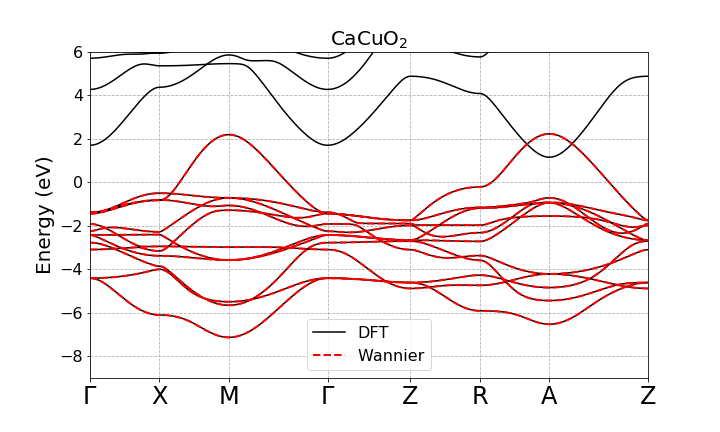}
    \caption{Wannier fits in the wide energy window.}
    \label{fig:wide_wannier}
\end{figure}

\begin{figure}[t]
    \centering
    \includegraphics[width = .44 \linewidth]{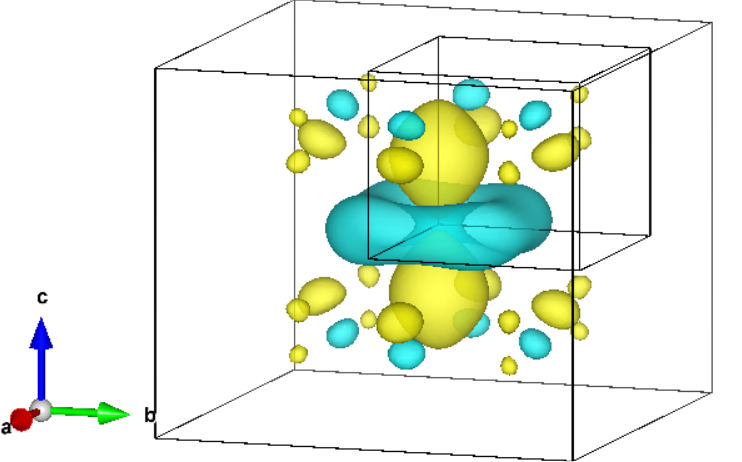}
    \includegraphics[width = .44 \linewidth]{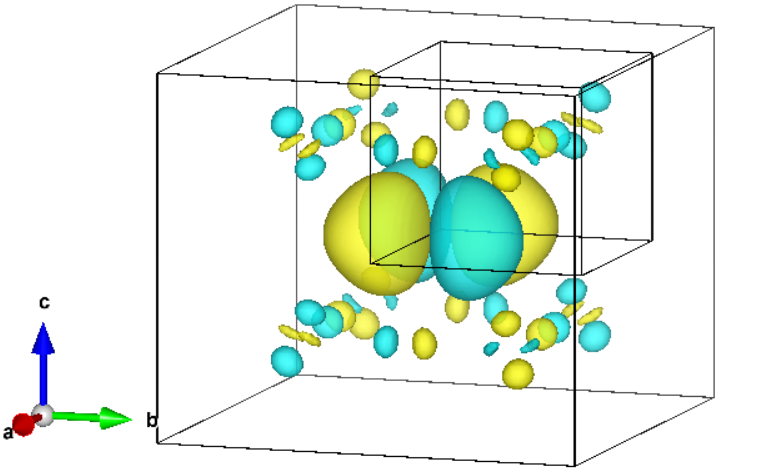}
    \caption{Real space plots of the narrow window Nd $d_{3z^2-r^2}$ (left) and $d_{xy}$ (right) Wannier functions.}
    \label{fig:nd_wanniers}
\end{figure}

\subsection{DMFT calculations}
We perform single-shot single-site DMFT calculations using the TRIQS software library~\cite{TRIQS, TRIQS/DFTTOOLS, TRIQS/CTHYB}. In the wide energy window calculations, we only treat the Ni/Cu-$e_g$ orbitals as correlated, but we include all of the Wannier functions in the DMFT self-consistency condition. In the narrow energy window calculation, we treat only the Ni/Cu-$d_{x^2-y^2}$ orbital as correlated, but in the nickelate case we also include the Nd orbitals in the self-consistency condition. For the DMFT calculations, we interpolate the Wannier function on a denser $40 \times 40 \times 40$ $k$-point mesh.

 For the impurity problem, we use a Kanamori interaction Hamiltonian~\cite{Kanamori1963} with $U' = U - 2J$. In the wide energy window calculation, we use an on-site Hubbard interaction of $U = \SI{7}{eV}$ and a Hund's coupling of $J = \SI{0.7}{eV}$ for both materials, generally accepted values~\cite{Nowadnick15, Eskes88}, at a temperature of $T = 290 K$. For the narrow energy window, we perform calculations at different values of $U$ and $T$ to construct a phase diagram. We allow for in-plane antiferromagnetism by doubling the unit cell in the $c(2 \times 2)$ scheme and forcing the self-energy for neighboring sites and opposite spins to have the same value. To solve the impurity problem, we employ the continuous-time hybridization expansion solver~\cite{TRIQS/CTHYB}. We typically use $\sim 10^8$ measurements, but we use $\sim 10^9$ when analytic continuation is needed and $\sim 10^7$ for parts of the phase diagram.  We employ Held's double counting formula~\cite{Held2007Electronic}:
 \begin{equation}
     \Sigma_{dc} = \frac{U + (D-1)(U-2J) + (D-1)(U-3J)}{2D-1} (n - 0.5),
 \end{equation}
  where $D$ is the number of correlated orbitals and $n$ is the total density, for which we take the density of the correlated orbitals obtained from their local non-interacting Green's function. We employ the maximum-entropy method~\cite{TRIQS/maxent} to perform analytic continuation on the Green's function and self-energy.
  
\subsection{Moving $\varepsilon_p$ down in the cuprate}
\begin{figure}[t]
    \centering
    \includegraphics[width = .48 \linewidth]{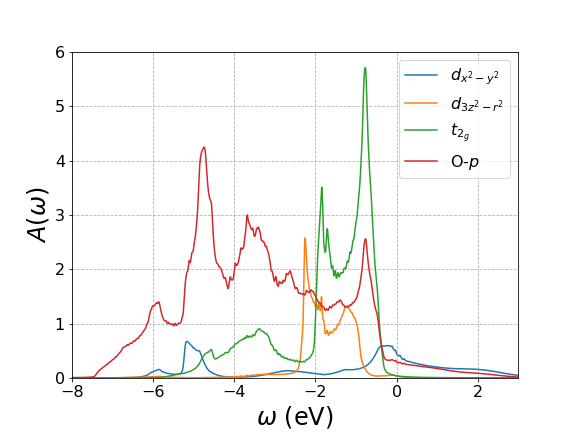}
    \caption{CaCuO$_2$ orbitally resolved spectral function in the wide energy window with $\varepsilon_p$ moved down by $\SI{1}{eV}$ and $U = \SI{7}{eV}$. Note that the chemical potential moves down by $\sim\SI{0.28}{eV}$ relative to the case where $\varepsilon_p$ is unchanged.}
    \label{fig:ep_down}
\end{figure}

We study the effect of lowering the O-$p$ orbitals in the wide energy window model of the cuprate. In the Hamiltonian matrix in the Wannier basis, we decrease the on-site energies of the O-$p$ orbitals by $\SI{1}{eV}$. As a result, the Cu-$d_{x^2-y^2}$ orbital hybridizes less with the O-$p$ orbitals. This makes the Cu-$d_{x^2-y^2}$ orbital more correlated in comparision to the calculation using the Held double counting, with a mass enhancement of $Z \approx 0.50$. Additionally, comparing Fig.~\ref{fig:ep_down} to main text Fig.~1, we see that the O-$p$ states have a smaller spectral weight at the chemical potential.

\subsection{Self-energies in wide and narrow window calculations}
Fig.~2 of the main text shows the imaginary part of the Matsubara self-energy for both materials in the orbital basis for both the wide and narrow energy windows. NdNiO$_2$ has roughly the same low energy behavior (and therefore the same mass enhancement) for the wide energy window with $U = \SI{7}{eV}$ as the narrow energy window with $U = \SI{3.1}{eV}$, indicating that $U \approx \SI{3}{eV}$ is a reasonable choice for the narrow energy window calculation. For CaCuO$_2$, in order to match the narrow energy window mass enhancement to the wide energy window one (with the Held double counting), we would need to use a $U$ of roughly $\SI{1.75}{eV}$. 

\begin{figure}[t]
    \centering
    \includegraphics[width = .48 \linewidth]{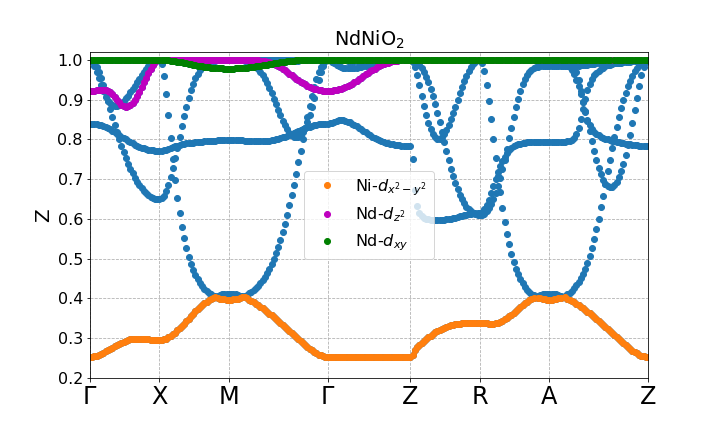}
    \includegraphics[width = .48 \linewidth]{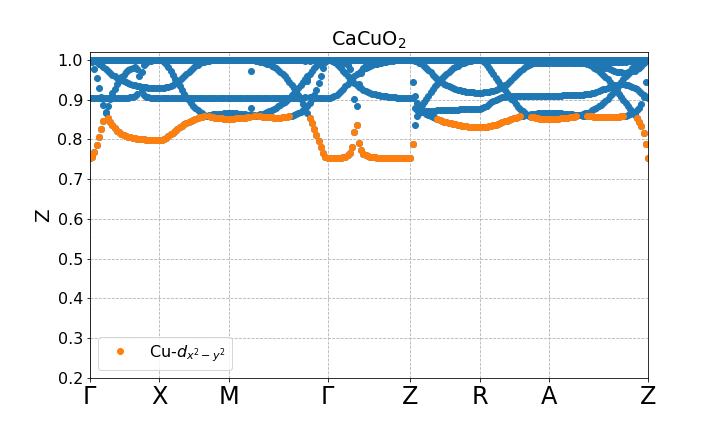}
    \caption{$Z$ for all states in the band basis for NdNiO$_2$ (left) and CaCuO$_2$ (right) in the wide energy window. The $Z$ values of bands with majority Ni/Cu-$d_{x^2-y^2}$ or Nd character are highlighted.}
    \label{fig:Zband}
\end{figure}

Fig.~\ref{fig:Zband} shows the values of $Z$ in the band basis as a function of $k$. As the hybridization with O-$p_\sigma$ increases, the $d_{x^2-y^2}$ bands become less correlated and $Z$ increases. The values of $Z$ for the Ni/Cu-$d_{x^2-y^2}$-derived band at the Fermi level are around $0.33$ for the nickelate and $0.83$ for the cuprate. 

\subsection{Identifying Phase Boundaries}
\begin{figure}[t]
    \centering
    \includegraphics[width = .45 \linewidth]{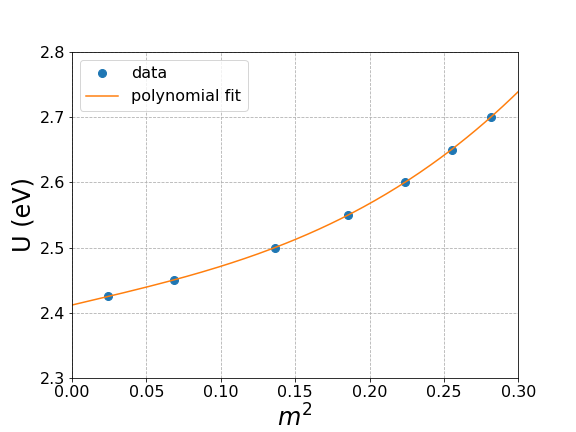}
    \caption{$U(m^2)$ for fixed temperature of $T = \SI{290}{K}$. We fit the points to a third order polynomial and identify the critical $U$ for the  phase transition to occur at about $U = 2.41$ in this example.}
    \label{fig:pt_example}
\end{figure}

For NdNiO$_2$, the paramagnetic to antiferromagnetic transition is found at fixed $U$ or $\beta$ by varying $\beta$ or $U$ just above the transition, fitting $\beta$ or $U$ as a function of magnetization squared to a polynomial, and taking the y-intercept of the resulting fit as the transition point. Fig.~\ref{fig:pt_example} shows an example at constant temperature. To find the metal-insulator transition in the nickelate, we make a linear fit to $\mu(U)$ in the region of the transition and choose the critical $U$ as the point where $\mu(U)$ equals the bottom of the DFT Nd band. For the cuprate where the transitions are sharper, we determine the transition points by bisection. Since the cuprate is at half filling, it is a metal if the Matsubara self-energy goes to $0$ as $\omega_n \to 0$ and an insulator if the Matsubara self-energy diverges as $\omega_n \to 0$. 

\begin{figure}[t]
\centering
\includegraphics[width= \columnwidth]{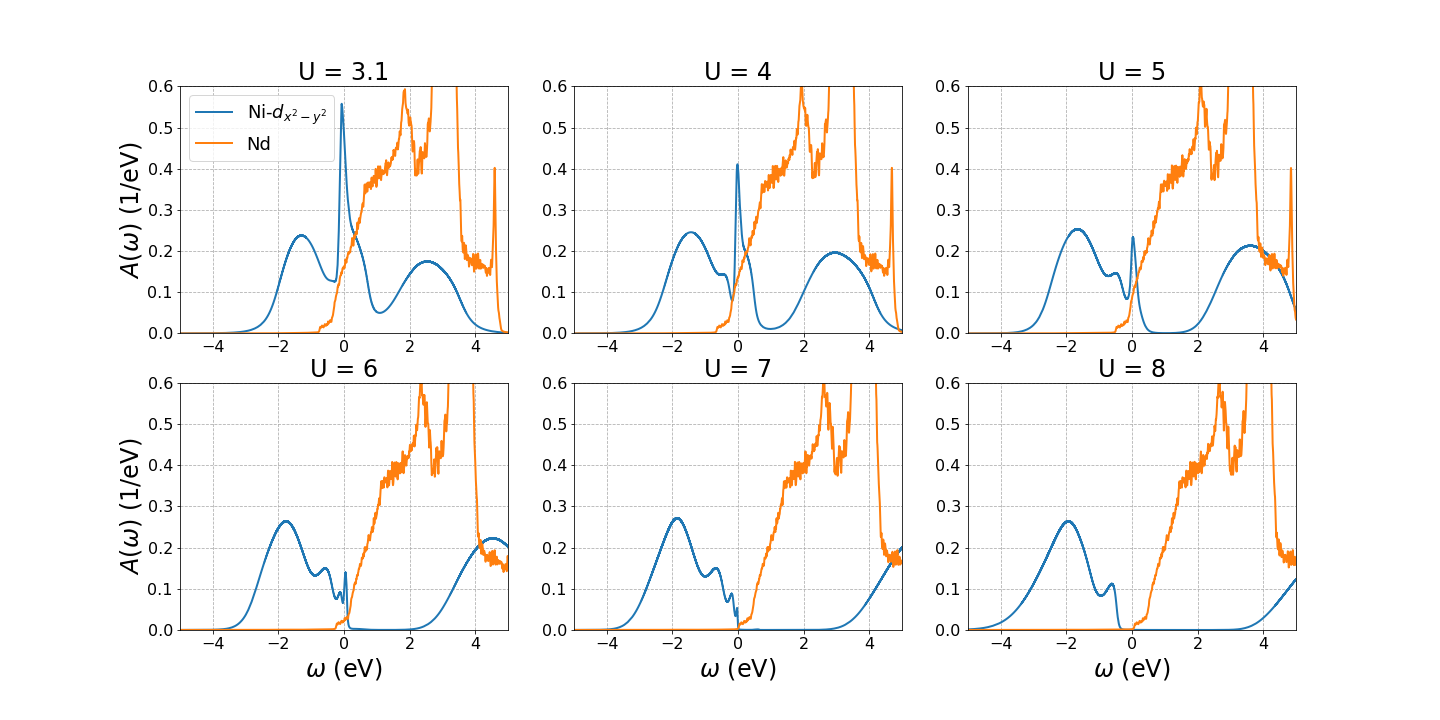}
\caption{NdNiO$_2$ narrow energy window paramagnetic spectral function of the Ni:$d_{x^2-y^2}$ (blue) and Nd (green) states. As $U$ increases, a gap opens in the $d_{x^2-y^2}$ DOS above the chemical potential. At around $U = \SI{7}{eV}$ the gap reaches the chemical potential when the Nd orbitals empty out.}
\label{fig:NNO_narrow_pdos}
\end{figure}

\begin{figure}[t]
\centering
\includegraphics[width=0.97\linewidth]{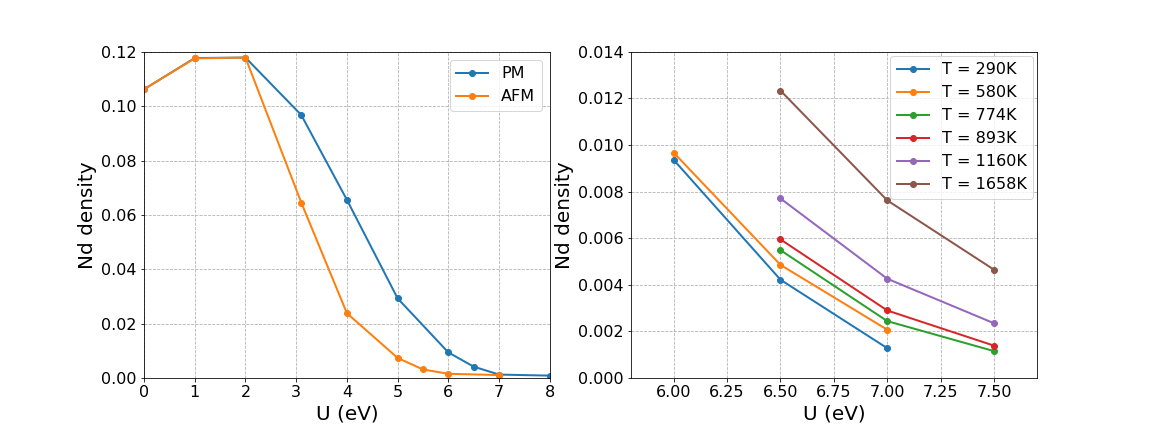}
\caption{Left panel shows the Nd density as a function of $U$ for the paramagnetic and antiferromagnetic cases at a fixed temperature of $T = \SI{290}{K}$. Right panel shows the Nd density as a function of $U$ in the paramagnetic case near the metal-insulator transition at various temperatures.}
\label{fig:nd_density}
\end{figure}

Fig.~\ref{fig:NNO_narrow_pdos} shows the spectral function in the paramagnetic phase for different values of $U$. The figure shows that the metal insulator transition coincides with the Nd bands emptying out. Fig.~\ref{fig:nd_density} shows that in the paramagnetic phase a higher value of $U$ is needed to empty out the Nd bands than in the antiferromagnetic phase, which explains why the metal-insulator transition is at a higher $U$ for the paramagnetic phase than the antiferromagnetic phase. The Nd occupancy is greater at higher temperatures, so the metallic phase is more stable at higher temperatures. 

\subsection{Role of Nd bands}
\begin{figure}[t]
    \centering
    \includegraphics[width = \linewidth]{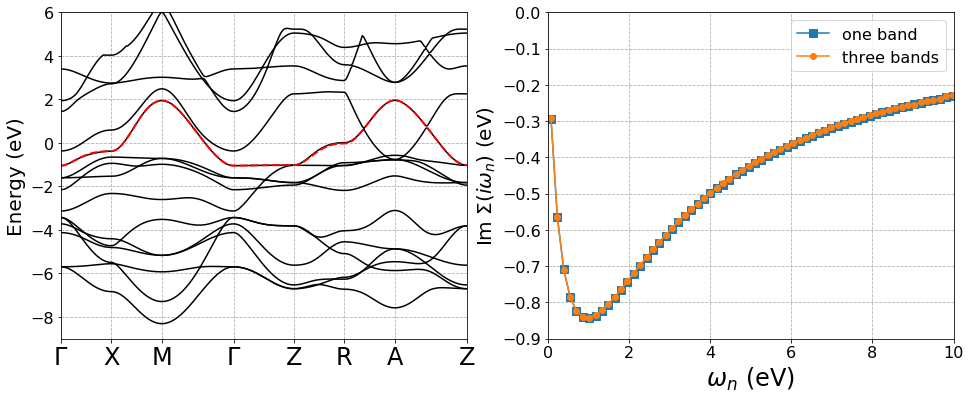}
    \caption{Left panel shows the one-band Wannier fit for NdNiO$_2$. Right panel compares the imaginary part of the Matsubara self-energy of the Ni-$d_{x^2-y^2}$ orbital for the one- and three-band models.}
    \label{fig:one_v_three}
\end{figure}

In the nickelate material in the narrow energy window calculation, the Nd orbitals hybridize only weakly with the Ni-$d_{x^2-y^2}$ state (see main text Tab.~2), but their main effect is to provide a doping reservoir for the system. We test this claim by doing a one band Wannier fit, keeping just the Ni-$d_{x^2-y^2}$-derived orbital. We fix the electron density to $0.903$, the value obtained in the three-band model with $U = \SI{3.1}{eV}$, and perform a DMFT calculation in the one-band model with the same $U = \SI{3.1}{eV}$. We find that the imaginary part of the Matsbuara self-energy is nearly indistinguishable from the three-band case (see Fig.~\ref{fig:one_v_three}). 

\newpage

\end{document}